\documentclass[singlecolumn]{aastex63}
\usepackage[utf8]{inputenc}
\usepackage{bm}
\usepackage{textcomp}
\usepackage{natbib}
\usepackage{graphicx}
\usepackage{amssymb}
\usepackage{amsmath}
\usepackage{gensymb}

\accepted{June 29, 2021}

\submitjournal{ApJ}
\shorttitle{The polarization of the Ly-$\alpha$ lines of \ion{H}{1} and \ion{He}{2} for probing 
	the solar corona}
\shortauthors{Supriya et al.}

\begin{document}
\title{The polarization of the Lyman-$\alpha$ lines of \ion{H}{1} and \ion{He}{2} as a tool for exploring the solar corona}
	
\correspondingauthor{Supriya Hebbur Dayananda}
\email{supriya@iac.es}

\author{Supriya Hebbur Dayananda}
\affiliation{Instituto de Astrofísica de Canarias, E-38205, La Laguna, Tenerife, Spain}
\affiliation{Departamento de Astrofísica, Facultad de Física, Universidad de La Laguna, Tenerife, Spain}

\author{Javier Trujillo Bueno}
\affiliation{Instituto de Astrofísica de Canarias, E-38205, La Laguna, Tenerife, Spain}
\affiliation{Departamento de Astrofísica, Facultad de Física, Universidad de La Laguna, Tenerife, Spain}
\affiliation{Consejo Superior de Investigaciones Científicas, Spain}

\author{Ángel de Vicente}
\affiliation{Instituto de Astrofísica de Canarias, E-38205, La Laguna, Tenerife, Spain}
\affiliation{Departamento de Astrofísica, Facultad de Física, Universidad de La Laguna, Tenerife, Spain}

\author{Tanausú del Pino Alemán}
\affiliation{Instituto de Astrofísica de Canarias, E-38205, La Laguna, Tenerife, Spain}
\affiliation{Departamento de Astrofísica, Facultad de Física, Universidad de La Laguna, Tenerife, Spain}

\begin{abstract}
	The near-Earth space weather is driven by 
	the quick release of magnetic free energy in the solar corona. 
	Probing this extremely hot and rarified region of the 
	extended solar atmosphere requires 
	modeling the polarization of forbidden and permitted coronal lines.  
	To this end, it is important to develop efficient codes to calculate the Stokes 
	profiles that emerge from given three-dimensional (3D) coronal models, 
	and this should be done  
	taking into account the symmetry breaking produced by the presence 
	of magnetic fields and non-radial solar wind velocities. We have developed 
	such a tool with the aim of theoretically predicting and interpreting 
	spectropolarimetric observations of the solar corona in permitted and forbidden lines.  
	In this paper we show the results of a theoretical investigation of the linear
	polarization signals produced by scattering processes 
	in the \ion{H}{1} Ly-$\alpha$ line at 1216~\AA\ 
	and in the \ion{He}{2} Ly-$\alpha$ line 
	at 304~\AA\, using 3D coronal models by Predictive Science Inc. 
	These spectral lines have very different critical magnetic fields 
	for the onset of the Hanle effect (53~G and 850~G, respectively), 
	as well as different sensitivities to the Doppler effect caused by the solar wind velocities. 
	We study under which circumstances simultaneous observations of 
	the scattering polarization in these Ly-$\alpha$ lines can 
	facilitate the determination of magnetic fields and macroscopic velocities in the solar corona.  
\end{abstract}


\section{Introduction} \label{sec:intro}

There are two types of spectral lines that encode 
information on the $10^6$ K plasma of the solar corona: 
forbidden lines at visible and infrared (IR) wavelengths 
and permitted lines in the ultraviolet (UV) spectral region. 
The polarization that some physical mechanisms 
introduce in such spectral lines is sensitive 
to the magnetic fields of the solar 
corona \citep[e.g., the reviews by][]{Casini+2017,TrujilloBueno+2017}. 
These physical mechanisms are the scattering of the 
anisotropic radiation coming from 
the underlying solar disk and the Hanle and Zeeman effects \citep[see][]{LL04}. 
The circular polarization signals are dominated 
by the Zeeman effect, but  
they are very hard or impossible to measure because 
their amplitudes ${\sim} {{{\lambda}B}\over{\sqrt{T}}}$,  
$\lambda$ being the spectral line wavelength and $T$ and $B$ 
the temperature and magnetic field strength of the coronal plasma, respectively. 
The linear polarization signals are caused by scattering processes, 
and in the presence of a magnetic field inclined with respect to the symmetry axis of the 
incident radiation field they are modified by the Hanle effect. The critical magnetic field 
strength ($B_H$) for the onset of the Hanle effect in a spectral line 
is inversely proportional to the lifetime of the line's upper level.
For the typical magnetic field strengths ($B$)   
expected in the solar corona, the forbidden lines are in the saturation regime of the
Hanle effect (i.e., their linear polarization is sensitive 
only to the magnetic field orientation, because 
$B{\gg}5B_H$), while the permitted lines are in the unsaturated regime 
(i.e., their linear polarization is sensitive to both the
orientation and the strength of the coronal magnetic field, because typically  
$0.2B_H\,{\lesssim}\,B\,{\lesssim}\,5B_H$). 
In the present paper, we focus on the linear polarization produced 
by scattering processes in two permitted lines of the
solar ultraviolet spectrum, which have very different sensitivities to the Hanle effect.

One of the spectral lines considered here 
is the hydrogen Ly-$\alpha$ line at 1216~\AA, whose critical magnetic 
field strength for the onset of the Hanle 
effect is $B_H{\approx}53$~G. This spectral line originates all through the 
upper chromosphere and it shows a 
broad intensity profile in emission with a small depression 
at its core \citep[e.g.,][]{Warren+1998}. The residual neutral hydrogen atoms in 
the solar corona scatter the intense Ly-$\alpha$ line radiation coming from the underlying 
chromosphere \citep{Gabriel+1971,Moses+2020}. 
At coronal heights, the anisotropy of the Ly-$\alpha$ emission line 
radiation coming from the solar chromosphere is substantial. 
Consequently, the scattered coronal 
Ly-$\alpha$ radiation is expected to be linearly polarized. 
In the idealized situation of a spherically symmetric and static solar 
atmosphere, the linear polarization of the scattered radiation is parallel to the solar limb. 
In the presence of a non-radial magnetic field in the (optically-thin) corona of such idealized solar atmosphere 
the radiation field coming from the underlying solar disk still has axial symmetry around the radial direction, but   
the Hanle effect caused by the inclined magnetic field modifies the linear polarization of the zero-field 
case \citep{Bommier+1982,1992SPIE.1546..402F}. However, the Hanle effect is not the only mechanism capable 
of modifying the linear polarization produced by scattering in permitted lines like 
hydrogen Ly-$\alpha$ \citep[e.g., chapter 12 of][and references therein]{LL04}. Of 
particular importance is the Doppler effect caused by the solar wind velocity, which has to be taken into 
account because at the UV wavelengths of the permitted lines the intensity of the solar disk radiation has 
spectral structure (e.g., an emission profile for the Ly-$\alpha$ line of \ion{H}{1}). 
At each point in the solar corona 
its rarefied plasma is moving with a macroscopic velocity, with the implication that the mean intensity ($J^0_0$) 
of the radiation field as seen in the comoving frame increases (decreases) with increasing velocity for the case 
of an incident radiation field with an absorption (emission) line. In addition to this well-known Doppler 
brightening (dimming) effect, the anisotropy of the incident radiation field in the comoving reference frame is 
modified depending on the modulus and inclination of the macroscopic velocity with respect to the 
solar radial direction. Moreover, at points in the solar corona where the macroscopic velocity of the 
plasma is non-radial, we may have a symmetry breaking because the radiation 
coming from different points of the underlying solar disk
with the same line of sight (LOS) inclination is differently affected by the Doppler 
effect \citep{Sahal-Brechot+1986,Sahal-Brechot+1998,Khan+2011,Khan-Landi2012}.

The other spectral line investigated here is the Ly-$\alpha$ line of \ion{He}{2} at 304~\AA, whose critical magnetic 
field for the onset of the Hanle effect is $B_H{\approx}850$~G \citep{2012ApJ...746L...9T}. 
This spectral line originates in the chromosphere-corona transition 
region (TR) and it shows an intensity profile in emission, 
which is narrower than that of the hydrogen Ly-$\alpha$ 
line \citep{Doschek+1974,Cushman+1975,Cushman-Rense1978}.  
There are residual \ion{He}{2} ions in the solar corona, 
which scatter the \ion{He}{2} 304~\AA\ line radiation coming from the underlying 
TR \citep{Gabriel+1995,Moses+2020}. At coronal heights the anisotropy of this spectral line radiation is 
substantial, and we therefore expect that the scattered 
coronal Ly-$\alpha$ line of \ion{He}{2} at 304~\AA\ is also linearly polarized. However, in contrast with the 
case of the hydrogen Ly-$\alpha$ line, we expect that the Ly-$\alpha$ line of \ion{He}{2} is rather insensitive 
to the weak magnetic fields of the solar corona (because $B_H{\approx}850$~G for the \ion{He}{2} line), but 
much more sensitive to the solar wind velocities (because of the much narrower \ion{He}{2} emission 
line radiation coming from the TR).     

In this paper we consider state-of-the-art three-dimensional (3D) models of the solar corona and calculate 
the maps of the linear polarization signals produced by scattering in the Ly-$\alpha$ lines of \ion{H}{1} and \ion{He}{2}, 
taking into account and neglecting the magnetic field and macroscopic velocity of the coronal models.  
Our aim is to quantitatively investigate whether the different sensitivities of this pair of Ly-$\alpha$ lines to 
the coronal magnetic field and to the solar wind outflows can be exploited for facilitating the diagnostics of 
these quantities. To the best of our knowledge, this is the first time that the scattering polarization of the 
Ly-$\alpha$ line of \ion{He}{2} produced by the residual \ion{He}{2} ions of solar coronal models is investigated.
The works of \cite{2012ApJ...746L...9T} and \cite{Belluzzi+2012} concerned 
the linear polarization of the Ly-$\alpha$ lines of \ion{H}{1} and \ion{He}{2} produced by scattering in 
the upper solar chromosphere and in the transition region, which required taking into account the effects of radiative transfer. 

In section \ref{sec:formulation} we describe the formulation of the problem. In
section \ref{sec:results-pos} we discuss some of the theoretical results,
emphasizing the importance of velocity fields in understanding the
    linear polarization signals of the Ly-$\alpha$ lines of \ion{H}{1} and
    \ion{He}{2} in the solar corona. Furthermore, in section
\ref{sec:results-3D} we describe the two 3D Predictive Science Inc. 
models we have chosen and the emergent Stokes profiles calculated 
taking into account both the velocity and magnetic field of the models under 
consideration. Finally, we present our conclusions in section \ref{sec:conclusion}. 

\section{Formulation of the problem} \label{sec:formulation}
A detailed review of the physics of the spectral line polarization that results from the 
resonance scattering of solar-disk photons by residual coronal atoms (e.g., of \ion{H}{1} or \ion{He}{2}) 
can be found in section 3 of \cite{TrujilloBueno+2017}. Here we summarize the main physical ingredients 
of the problem's formulation.
Both Ly-$\alpha$ lines result from the transitions between levels $n=1$ and $n=2$, with $n$ the principal 
quantum number. The $n=1$ ground level is composed of the singlet $1s^2{S}_{1/2}$, while the upper 
level $n=2$ consists of the singlet $2s^2{S}_{1/2}$ and the doublet $2p^2{P}_{1/2,3/2}$. In both Ly-$\alpha$ lines, 
the radiation observed in the solar corona is dominated by resonance 
fluorescence \citep{Gabriel+1971,1997SoPh..175..645R,Patchett+81,Gabriel+1995,Moses+2020}. 
Since collisional processes are expected to 
be negligible in the formation of these Ly-$\alpha$ lines in the solar 
corona, the upper level $2s^2{S}_{1/2}$
cannot be populated and it can be ignored\footnote{We are also assuming that 
electric fields in the solar corona do not play any role on the 
excitation of this level \citep[see,][for information on the 
possible impact of an electric field]{1987A&A...179..329F, 2005PhRvA..71f2505C}.}. 
The fact that the separation between 
the $2p^2{P}_{1/2}$ and $2p^2{P}_{3/2}$ fine structure (FS) upper levels is 
about 17 times larger than their natural width, imply that in a weakly magnetized and 
optically thin medium like the solar corona we can safely neglect 
any quantum mechanical interference between the magnetic sublevels pertaining to such two
upper levels. Both Ly-$\alpha$ lines result from two transitions 
between the $1s^2{S}_{1/2}$ lower level and the $2p^2{P}_{1/2}$ and $2p^2{P}_{3/2}$ upper levels, 
which are blended because their separation is much smaller than the line's Doppler width.   
Although the only level that contributes to the scattering polarization 
in these Ly-$\alpha$ lines
is the upper level with angular momentum $J=3/2$, 
the FS of the Ly-$\alpha$ lines must however 
be taken into account because it reduces the polarizability factor of the lines. 
Fortunately, as shown by \cite{Bommier+1982}, 
the hyperfine structure can be safely neglected for modeling the scattering polarization
of Ly-$\alpha$ in the solar corona. 
In summary, a reliable modeling of the Ly-$\alpha$ radiation 
scattered by the solar corona can be achieved 
by means of a three-level model atom with the $1s^2{S}_{1/2}$
ground level and the $2p^2{P}_{1/2}$ and $2p^2{P}_{3/2}$ 
upper levels, without any quantum interference between 
them. 

The linear polarization produced by the scattering of anisotropic radiation 
in both Ly-$\alpha$ lines is sensitive to magnetic fields with strengths  
between $0.2B_{\rm H}$ and $5B_{\rm H}$, approximately, where 
$B_{\rm H}=1.137{\times}10^{-7}/(t_{\rm life}g)$ is 
the critical magnetic strength for the onset of 
the Hanle effect (with $t_{\rm life}$ and $g=4/3$ the radiative lifetime in seconds 
and the Land\'e factor of the line's upper level
with $J=3/2$, respectively). 
While $B_{\rm H}{\approx}53$~G for the hydrogen Ly-$\alpha$ line at 1216~\AA, it is 
$B_{\rm H}{\approx}850$~G for the Ly-$\alpha$ 
line of \ion{He}{2} at 304~\AA\ \citep{2012ApJ...746L...9T}. 
This is because $t_{\rm life}\,{\approx}\,1/A_{ul}$, and the Einstein coefficient $A_{ul}$ 
for spontaneous emission from the upper ($u$) to the lower ($l$) level for the 
Ly-$\alpha$ line of \ion{H}{1} at 1216~\AA\ 
is $A_{ul}\,{\approx}\,6.264{\times}10^8\,{\rm s}^{-1}$ 
 and is $A_{ul}\,{\approx}\,1.0029{\times}10^{10}\,{\rm s}^{-1}$ 
for the \ion{He}{2} line at 304~\AA, which is 16 times larger than the former.

To model the polarization of solar coronal lines, we have developed a computer code based on the multilevel atom 
theory described in section 7.2 of \cite{LL04}. Our choice for the quantization axis for total angular momentum is 
the solar radius vector through the considered spatial point in the solar corona; therefore,  
the statistical equilibrium equations are equations (7.78) given in such monograph, which 
can be directly solved once the incident radiation field is specified for each radiative transition in the 
multilevel model under consideration. The numerical 
solution of these equations gives the multipolar components $\rho^K_Q(J)$ of the atomic density matrix 
for each atomic level of total angular momentum $J$. 
Since the solar corona is optically thin at the wavelengths of the 
considered Ly-$\alpha$ lines, we only have to calculate 
the emission coefficient $\epsilon_i (\nu, \bm \Omega)$ (where the index `$i$' take the values 
0,1,2, and 3, corresponding to Stokes $I$, $Q$, $U$, and $V$, respectively) 
for each line frequency $\nu$ and propagation  
direction $\bm \Omega$. This emission coefficient in the four 
Stokes parameters $I_i (\nu, \bm \Omega)$ has to be calculated at each position along the considered 
off-limb LOS, and we finally obtain the frequency-integrated Stokes signal 

\begin{equation}
I_i (\bm \Omega) = \int \mathrm{d}\nu\,\int_{LOS} \epsilon_i (\nu, \bm \Omega) \, \mathrm{d}s,
\label{Eqn-LOS}
\end{equation}
where $s$ is the geometrical distance along the LOS. 

The emission coefficient depends on the multipolar components $\rho^K_Q(J_u)$ of the atomic density matrix  
of the upper levels of the Ly-$\alpha$ lines, the calculation of which requires solving the statistical equilibrium 
equations for the three-level model atom mentioned above. 
Given that $J$-state interference do not play any role
here and that collisional processes are negligible, the two upper levels are not coupled and one is basically 
left with a pair of two-level atom equations for the two blended transitions. The crucial quantity that enters  
these equations is the radiation field tensor $J^K_Q$, which quantifies the symmetry properties 
of the Ly-$\alpha$ radiation that illuminates the \ion{H}{1} and \ion{He}{2} atoms of the solar corona:

\begin{eqnarray}
&& \!\!\!\!\!\!\!\!\!\!\!J^K_Q (\nu_0) = \int^\infty_\infty f({\bm v}-{\bm w})\, {\mathrm d}^3{\bm v}  \nonumber \\&&
\times \oint \frac{{\mathrm d}{\bm \Omega}^\prime}{4\pi} \, T^K_Q(0,{\bm \Omega}^\prime) \, I\Big(\nu_0\Big( 1+\frac{{\bm v.\bm \Omega}^\prime}{c}\Big),\bm \Omega^\prime\Big),
\label{Eqn-JKQ}
\end{eqnarray}
where $\nu_0$ is the transition frequency, $\bm v$ is the vector sum of the thermal velocity and of the solar 
wind velocity, $\bm w$, and $ f({\bm v}-{\bm w})$ is the velocity distribution function of the atoms in the 
corona. $T^K_Q(0,{\bm \Omega}^\prime)$ is the polarization tensor, which depends on the propagation 
direction ${\bm \Omega}^\prime (\theta^\prime,\chi^\prime)$ of the incoming radiation 
(see Figure 12.10 in LL04; $\Omega (\theta,\chi)$ in this figure 
corresponds to $\Omega^\prime (\theta^\prime,\chi^\prime)$). 
The last term in Eq. (\ref{Eqn-JKQ}) represents the Stokes-$I$ emission profile 
of the incoming Ly-$\alpha$ radiation, which is Doppler shifted as seen by the coronal atoms. 
This equation accounts for the loss of axial symmetry of the radiation field at any 
given point in the solar corona, due to the Doppler effect caused by the 
macroscopic velocity of the solar wind. 

In Eq. (\ref{Eqn-JKQ}) the term of the Doppler effect is 
\begin{eqnarray}
	&& {\bm v.\bm \Omega}^\prime = v \, [{\rm cos} \theta^\prime \, {\rm cos} \theta_v 
	+ {\rm sin} \theta^\prime \, {\rm sin} \theta_v \, {\rm cos} (\chi^\prime-\chi_v)],
\end{eqnarray}   
where $v$ is the modulus of the velocity and the $\theta_v$ and $\chi_v$ angles indicate its direction.
In a reference frame at rest with respect to the solar surface and having the 
Z-axis directed along the solar radius vector through the considered point in the solar corona, 
the radiation coming from the underlying solar disk is cylindrically symmetrical (i.e., independent of the azimuth), 
which implies that the only non-zero components of the radiation field tensor 
are $J^0_0$ (mean intensity) and $J^2_0$ (anisotropy). However, if the solar wind velocity 
has non-radial components, an observer 
in the comoving frame (i.e., the reference system moving with velocity $\bm v$ 
at the point under consideration) will see a radiation intensity that depends on the azimuth, 
which implies that also the $J^2_1$ and $J^2_2$ components will be non-zero. 
Clearly, in the case of a purely radial solar wind, $J^2_1=J^2_2=0$. 
However, $J^0_0$ and $J^2_0$ are modified with respect to the static case 
because, due to the Doppler effect, the coronal atoms experience a lower radiation 
intensity \citep[correspondingly the coronal atoms would experience Doppler-brightening 
if the incoming radiation were an absorption instead of an emission 
profile, see chapter 12 of][for more details]{LL04}.

\begin{figure*}
	\centering
	\includegraphics[scale=0.55]{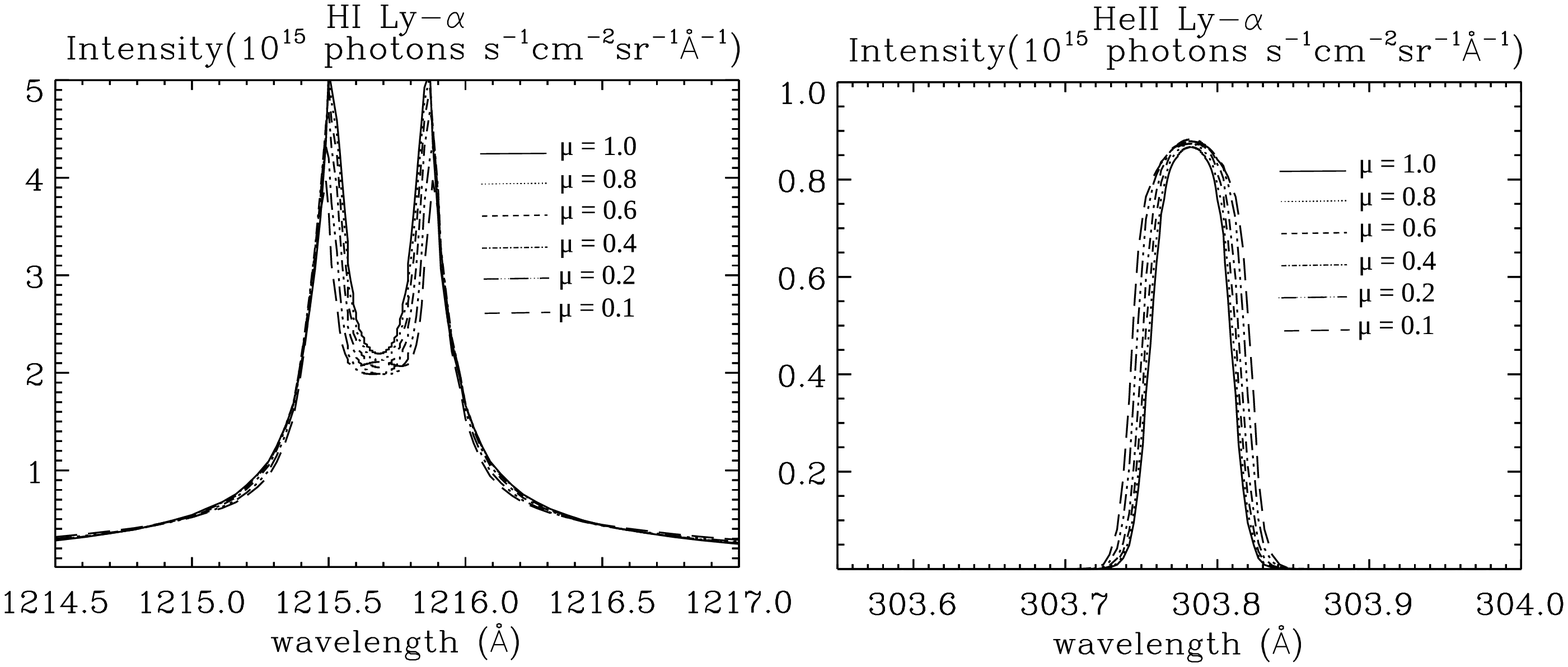}
	\caption{Center to limb variation of the intensity profiles of the 
	Ly $\alpha$ lines of H {\sc i} (left panel) and He {\sc ii} (right panel), 
	calculated in the FAL-C solar semi-empirical model.    
	The LOS is characterized by $\mu=cos\theta$, with $\theta$ the heliocentric angle.}
	\label{HIHeIICLV}
\end{figure*}

The following section shows some illustrative results useful to clarify how 
the various components of the (comoving frame) radiation field tensor 
of the Ly-$\alpha$ lines of \ion{H}{1} and \ion{He}{2} react to radial and non-radial solar wind
macroscopic motions. These academic results will be helpful in better understanding 
the results of 
Section 4, where we show the maps of the linear polarization signals in both Ly-$\alpha$ lines 
produced by scattering in 3D solar coronal models from Predictive Science Inc. In the calculations of this 
paper, the intensity profiles of the corresponding Ly-$\alpha$ line radiation that emerges from the 
quiet solar disk are given in Figure \ref{HIHeIICLV}. These Stokes-$I$ profiles result from non-equilibrium 
radiative transfer calculations with the RH code of \citet{2001ApJ...557..389U} 
in the semi-empirical model C of \cite[][hereafter, FAL-C]{Fontenla+1993}. 
The radiation at the \ion{H}{1}
Ly-$\alpha$ line is calculated for a 10 level (9 \ion{H}{1} levels + \ion{H}{2}) model atom without fine structure,
taking into account partial frequency redistribution effects in both Ly-$\alpha$ and Ly-$\beta$ transitions. The
radiation at the \ion{He}{2} Ly-$\alpha$ line is calculated for a 53 level (46 \ion{He}{1} levels +
6 \ion{He}{2} levels + \ion{He}{3}) model atom, with fine structure only in the \ion{He}{1} atomic levels and
taking into account partial frequency redistribution effects in the \ion{He}{2} Ly-$\alpha$ line. Both atomic models
are part of the RH code, but we have modified them to increase the number of frequency nodes (in particular, the
relevant Ly-$\alpha$ transitions are sampled with 200 frequency nodes) and to account for partial redistribution
effects in the \ion{He}{2} Ly-$\alpha$ line.
As seen in Figure \ref{HIHeIICLV}, the center-to-limb variation of these intensity profiles is not very significant.
At coronal heights the anisotropy of the spectral line radiation in both Ly-$\alpha$ lines is fully dominated 
by the fact that the larger the height above the visible solar disk sphere the
smaller the solid angle subtended.    
We have accounted for this fact as explained in Section 12.3 of \cite{LL04}. 

\section{Doppler dimming and symmetry breaking} \label{sec:results-pos}

It is of interest to start illustrating the impact of radial and non-radial solar wind velocities 
on the comoving frame radiation field tensor \citep[cf.,][]{LL04}. To this end, we assume 
an $10^6$ K isothermal solar coronal model with a constant outward macroscopic velocity    
and calculate the components of the radiation field tensor for 
the Ly-$\alpha$ lines of \ion{H}{1} and \ion{He}{2}. As shown in Figure \ref{HIHeIICLV}, the 
intensity profiles of the Ly-$\alpha$ line radiation coming from the underlying solar disk have very different widths, 
with the \ion{H}{1} 1216~\AA\ line being $\sim$8.5 times broader than the \ion{He}{2} 304~\AA\ line.

\begin{figure*}
    \centering
    \includegraphics[scale=0.7]{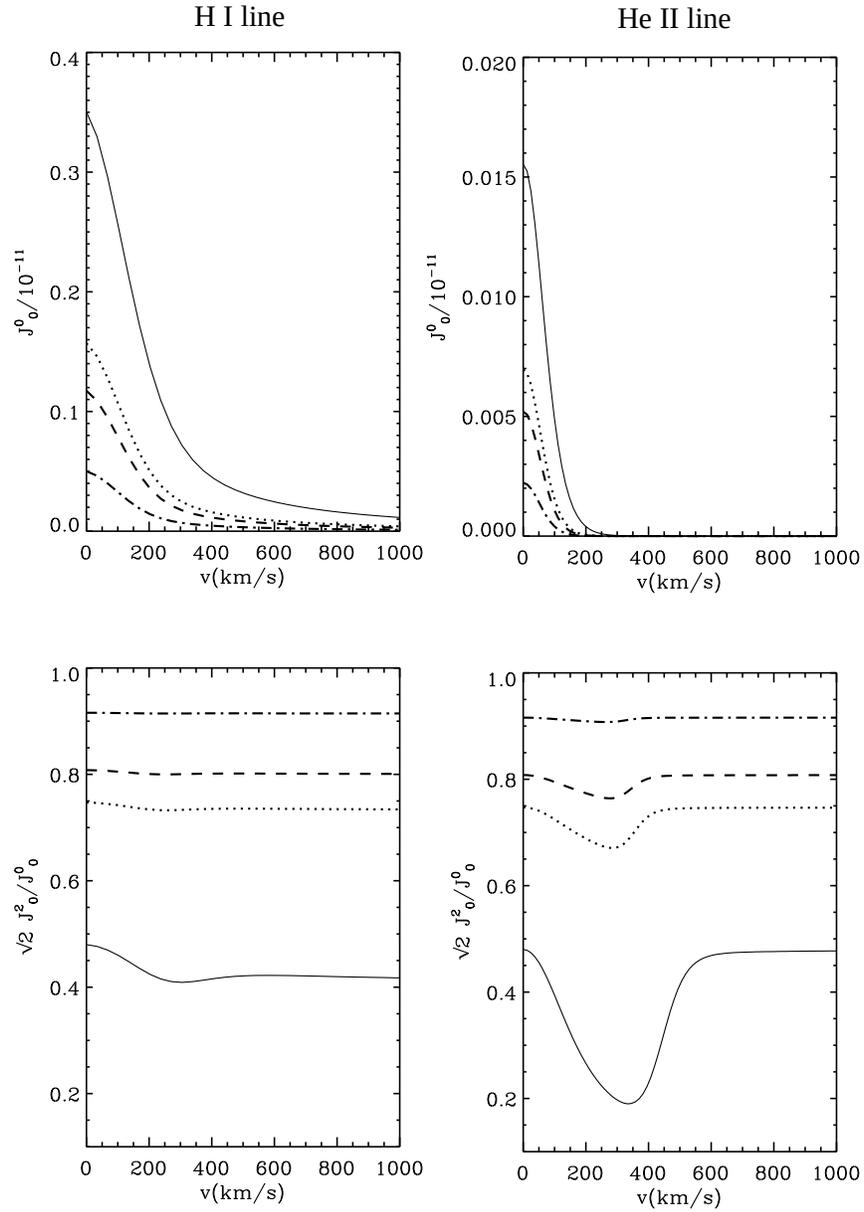}
    \caption{Variation of the indicated components of the comoving frame 
    radiation field tensor with the velocity of the assumed radial solar wind, 
    for the H {\sc i} Ly-$\alpha$ line (left panels) and the He {\sc
      ii} Ly-$\alpha$ line (right panels). The different curves correspond to the following heights above the solar visible disk: 
    solid h=0.25R$_\odot$, dotted h=0.75R$_\odot$, dashed h=1.0R$_\odot$, and dot-dashed h=2.0R$_\odot$. 
    }
    \label{JKQHIHEII-radial}
\end{figure*}

First, we consider a range of solar wind velocities (0-1000 km/s) directed along the solar radius and 
calculate the radiation field tensor at various heights above the Sun's visible disk. 
Given that the macroscopic velocity is radial and that the underlying solar disk is 
assumed to be devoid of any structure (e.g., sunspots) capable of breaking the axial 
symmetry of the incident radiation field, the only non-zero components of the radiation field 
tensor are $J^0_0$ and $J^2_0$. Figure \ref{JKQHIHEII-radial} shows the results 
for the Ly-$\alpha$ lines of \ion{H}{1} (left panels) and \ion{He}{2} (right panels), namely 
$J^0_0$ (upper panels) and $\sqrt{2}J^2_0/J^0_0$ (bottom panels) against the modulus of the radial solar 
wind velocity. We recall that the anisotropy factor of the radiation field at the spatial point under 
consideration is $\sqrt{2}J^2_0/J^0_0$, which is unity for 
the case of a unidirectional unpolarized radiation beam.

As expected, 
the mean intensity $J^0_0$ is larger for the \ion{H}{1} line, because for this spectral line 
the intensity that emerges from the underlying solar disk  
is about one order of magnitude larger than that corresponding to the \ion{He}{2} line. 
Also, the larger the height above the solar visible disk the smaller the mean intensity, because the  
solid angle subtended by the Sun's visible sphere decreases. The most 
noteworthy point is the rapid decrease of $J^0_0$ as 
the macroscopic velocity of the assumed radial solar wind 
increases. This so-called Doppler dimming effect occurs 
because, due to the Doppler effect, the emission peak 
of the radiation that comes from the underlying chromosphere and transition region, 
as seen from the coronal atoms, falls out of resonance. As a result, for sufficiently large velocities 
the solar wind atoms see, at frequency $\nu_0$, the radiation emitted by the underlying 
solar disk in the far wings of the line, where the intensity of the Ly-$\alpha$ line under consideration 
is smaller than at the line center. As expected, the Ly-$\alpha$ line of \ion{He}{2} is more sensitive to 
this Doppler dimming effect, because the intensity profile of the incoming radiation is narrower.

As seen in the bottom panels of 
Figure \ref{JKQHIHEII-radial}, in the static case, the anisotropy factor of both Ly-$\alpha$ lines 
is exactly the same, simply because the two lines have the same polarizability factor. When the 
solar coronal plasma is moving radially the anisotropy factor first decreases and then grows, 
as the magnitude of the velocity increases. The decrease occurs because the Doppler effect acts 
as a limb brightening: the nearly radial beams of the incoming radiation are 
redshifted and hence the coronal atoms see less intensity at frequency $\nu_0$,  
while the same effect is weaker for the predominantly horizontal radiation beams. Clearly, as soon as 
the velocity is sufficiently large the anisotropy factor tends to its static value because the 
coronal atoms see, at frequency $\nu_0$, the radiation emitted by the solar disk in the far wings of the line. 
This occurs much sooner for the narrower Ly-$\alpha$ line of \ion{He}{2}; therefore, the anisotropy factor in both 
lines is not exactly the same for 1000 km/s. 
Also, the sensitivity of the comoving frame 
anisotropy factor to the radial velocity is larger for the Ly-$\alpha$ line 
of \ion{He}{2}, because the intensity profile of its incoming radiation is narrower.

\begin{figure*}
	\centering
	\includegraphics[scale=0.5]{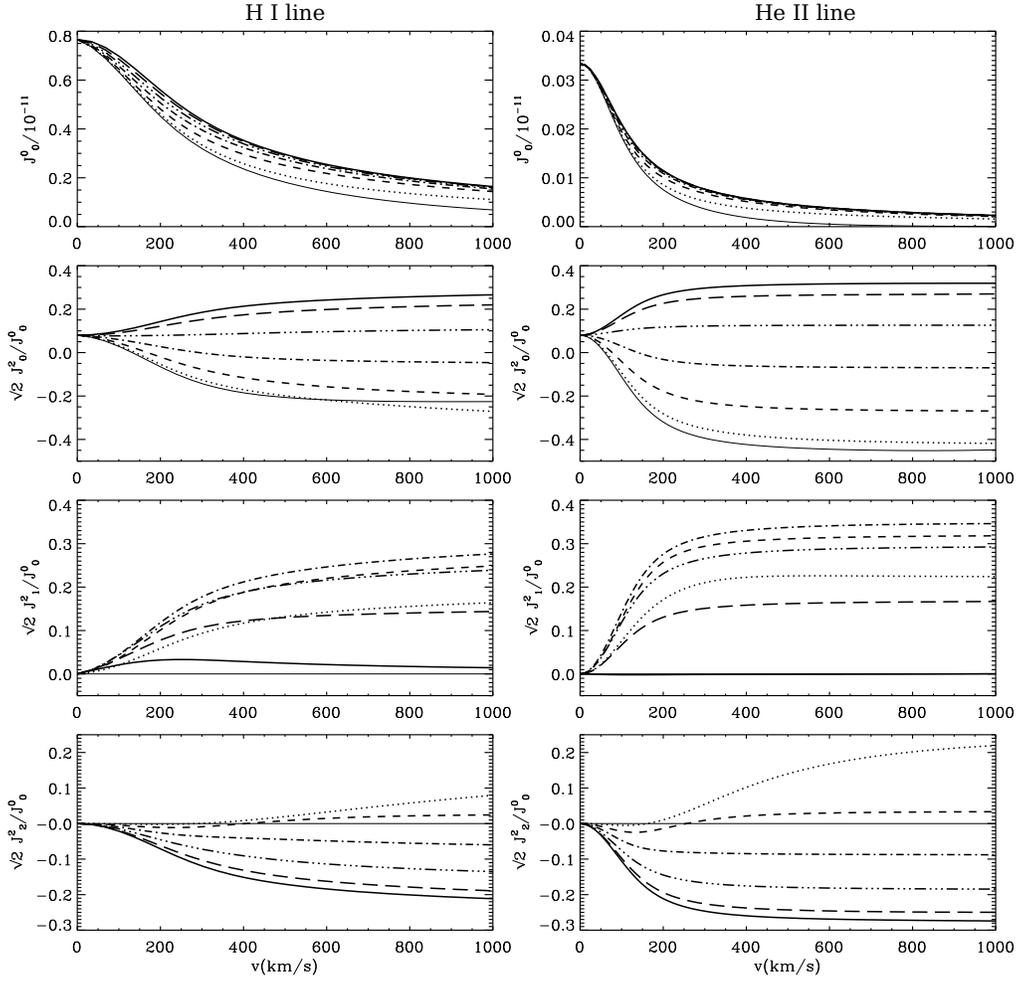}
	\caption{The impact of non-radial solar wind velocities on the radiation field tensors for the H {\sc i} 
	Ly-$\alpha$ line (left panels) and the He {\sc ii} Ly-$\alpha$ line (right panels) at a height $h=0.01{\rm R}_\odot$ above the solar surface. 
	The various curves are for the following inclinations of macroscopic velocity: thin solid-0$\degree$; dotted-15$\degree$; 
	dashed-30$\degree$; dot-dashed-45$\degree$; dashed triple-dotted-60$\degree$; long dashed-75$\degree$; thick solid-90$\degree$. 
	For all curves the azimuth angle of the velocity is 0$\degree$. We show only the real components of $J^2_1$ and $J^2_2$, 
	because their imaginary components are very small for the chosen azimuth.}
	\label{JKQHIHEII-nonradial-A}
\end{figure*}

The non-radial solar wind case for various inclinations of the macroscopic velocity
is shown in Figure \ref{JKQHIHEII-nonradial-A} 
for a height $h=0.01{\rm R}_\odot$ above the Sun's visible disk. It is of interest 
to note in the four top panels the behaviour of $J^0_0$ and $\sqrt{2}J^2_0/J^0_0$ for non-radial solar winds, 
which can be understood by arguments similar to those outlined above. For example, 
for the case of a solar wind velocity perpendicular to the radial direction, the maximum Doppler shift and the 
corresponding decrease in the intensity seen by the coronal atoms occurs for the radiation beams 
coming from the sides and directed along the velocity vector. 
Accordingly, the comoving frame anisotropy factor decreases for the inclination angles 0$\degree$ - 45$\degree$ and increases for the later angles. However they all 
reach the static value in the limiting case of infinite velocity which is much greater than 
1000 km/s for the representative case in Figure \ref{JKQHIHEII-nonradial-A}.
 Similar arguments can be made 
to understand the behaviour of the $J^2_1$ and $J^2_2 $ components, which 
quantify the breaking of the cylindrical symmetry of 
the radiation field experienced by the moving coronal atoms. 
The real components of $J^2_1$ and $J^2_2 $ 
are shown in Figure \ref{JKQHIHEII-nonradial-A}. 
As expected, because of the fulfilment of the cylindrical symmetry $J^2_1=J^2_2 =0$ 
for v=0, and also for the limit of very large velocities ($\gg$ 1000 km/s 
for the case considered 
in Fig. \ref{JKQHIHEII-nonradial-A}; for such large velocities the coronal atoms 
see the far wings of the line radiation coming from the solar disk). 

\section{Results in 3D coronal models from Predictive Science Inc.} \label{sec:results-3D}

In this section we show maps of the frequency-integrated 
linear polarization signals produced by scattering processes 
in the Ly-$\alpha$ lines of \ion{H}{1} and \ion{He}{2}, which we have calculated in 3D models of the solar corona.
Our aim is to show the different sensitivity of these lines to the model's magnetic field and macroscopic velocity.  
  
\subsection{The 3D models}  
We use 3D magneto-hydrodynamic models of the solar corona and inner heliosphere 
developed by Predictive Science Inc. (see https://www.predsci.com/portal/home.php). 
These publicly available coronal models are developed using photospheric magnetic field observations 
to specify the boundary condition on the radial component of the magnetic field, and they include 
energy transport processes such as coronal heating, anisotropic thermal 
conduction, and radiative losses \citep{2009ApJ...690..902L, 2015ApJ...802..105R}. 
These 3D models provide the proton number densities, the temperature, the magnetic field, 
and the velocity of the coronal plasma at each spatial point (in
spherical coordinates).

\begin{figure*}
	\includegraphics[scale=0.5]{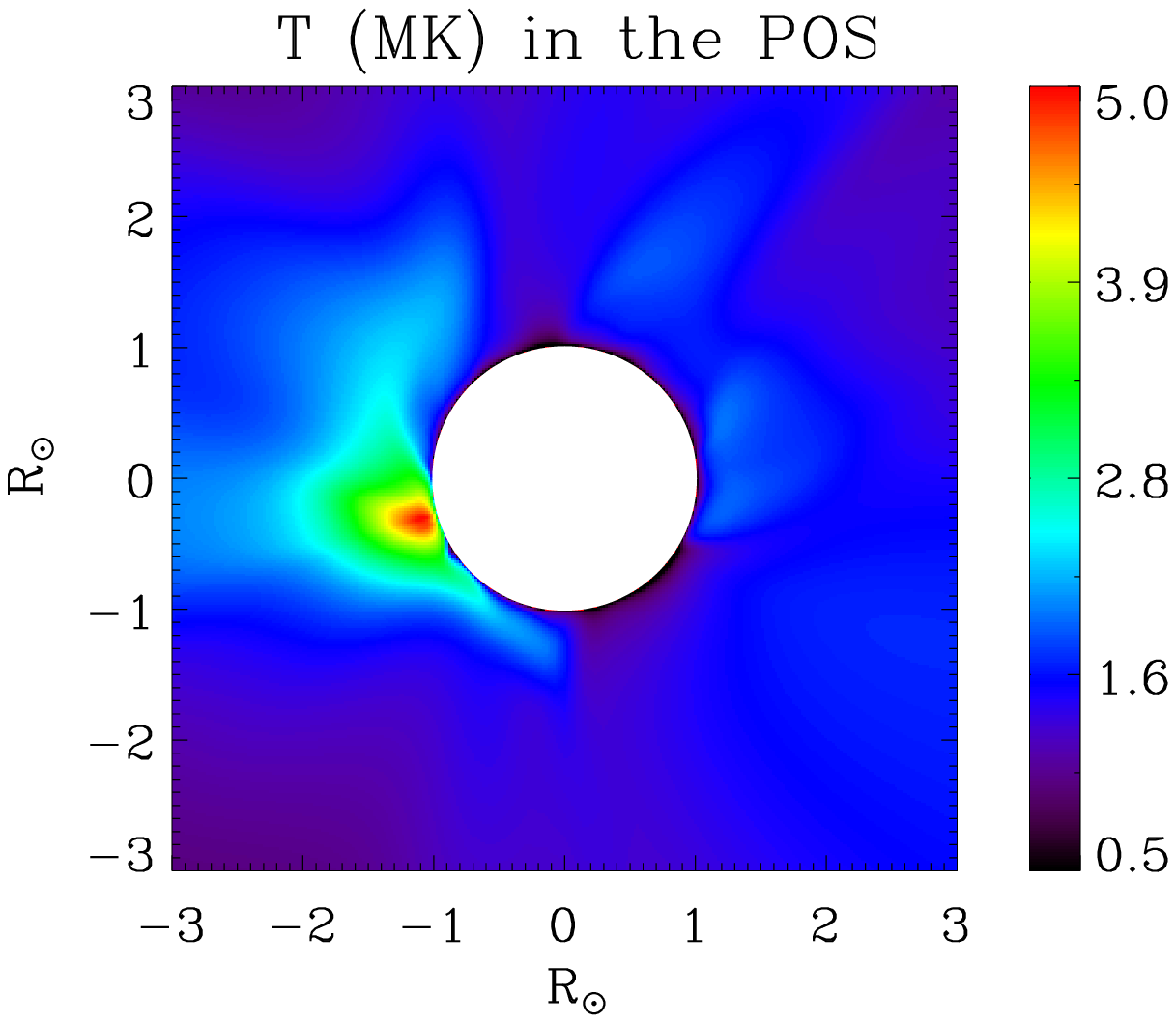}
	\includegraphics[scale=0.5]{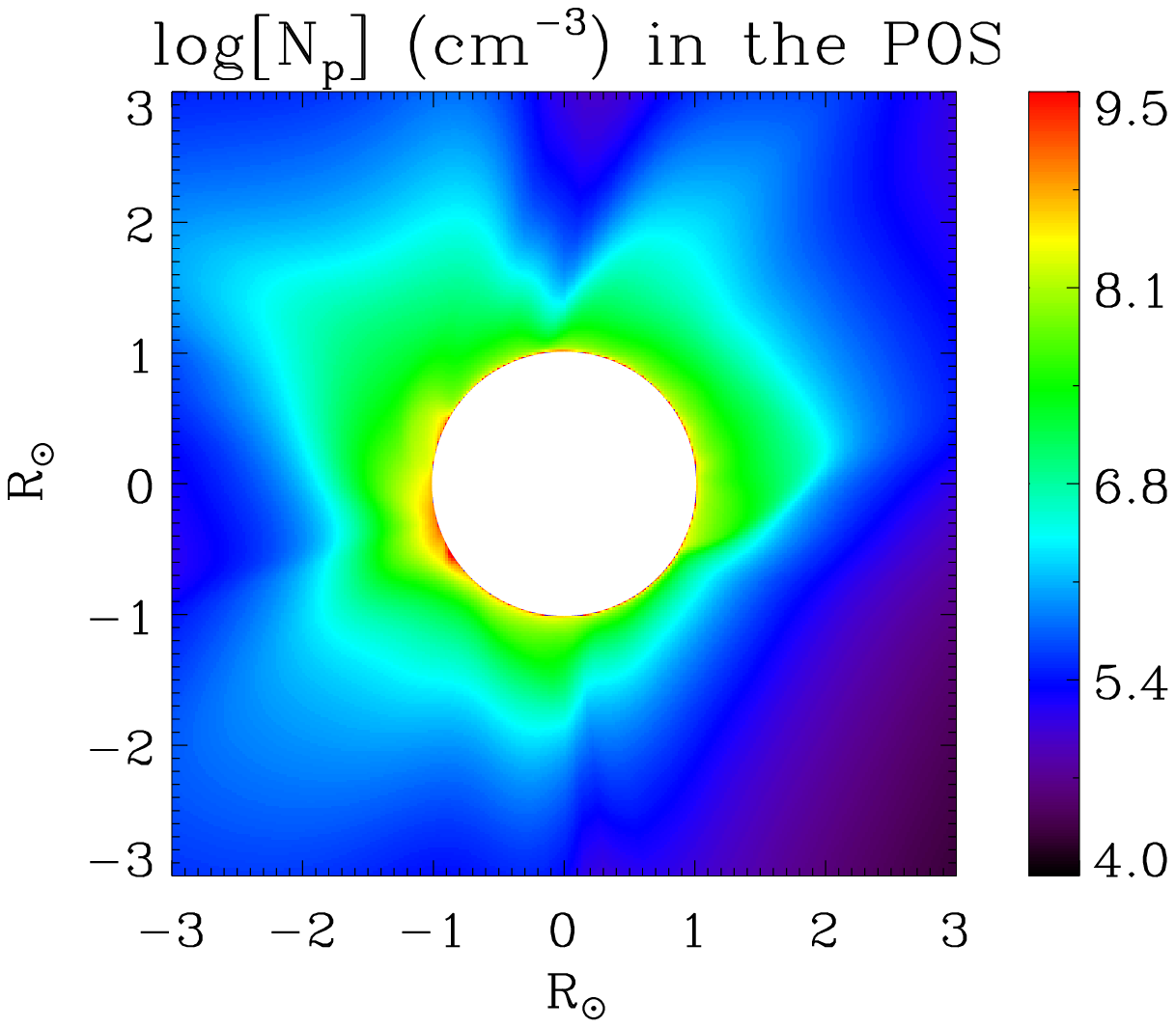}
	\includegraphics[scale=0.5]{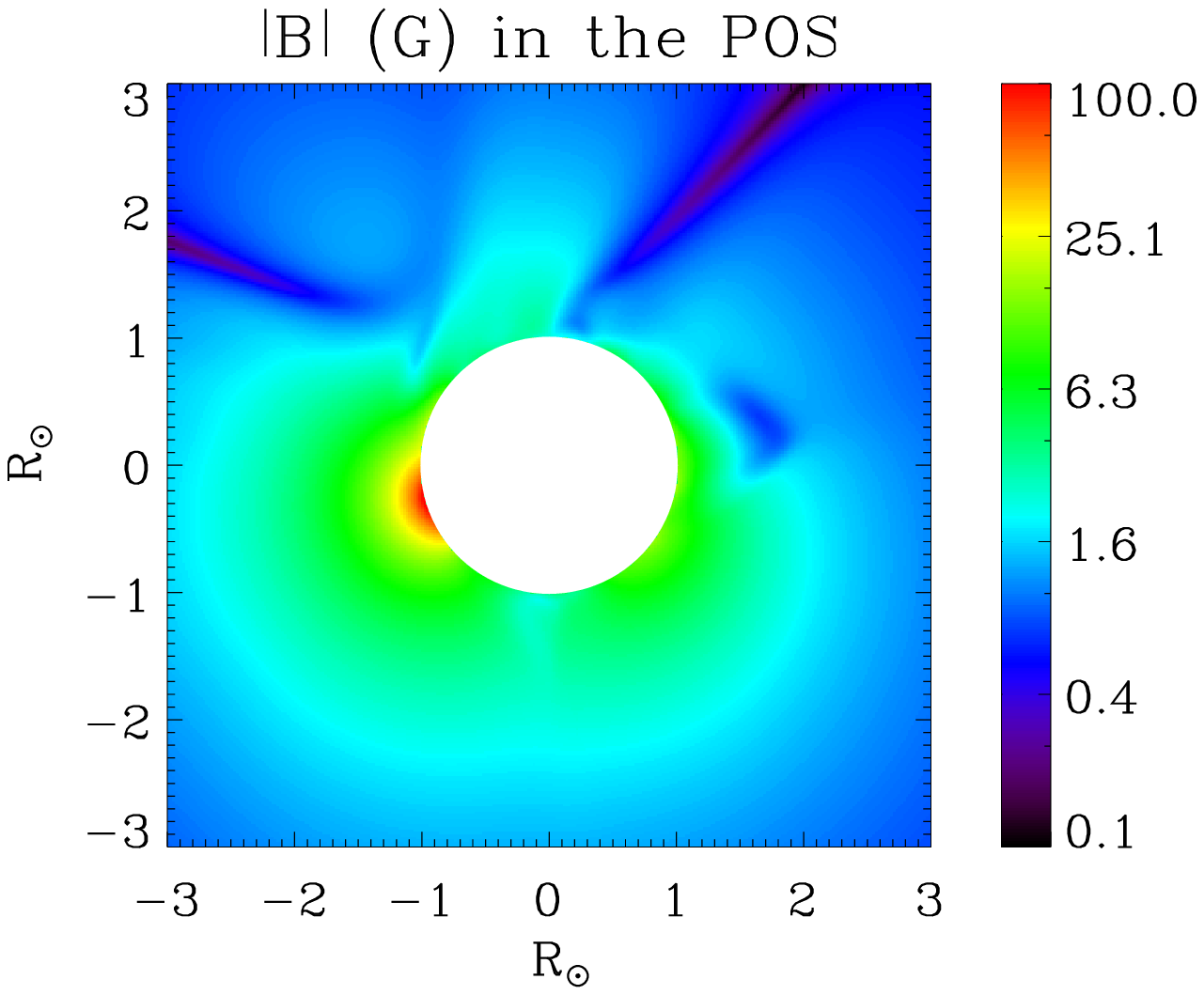}
	\includegraphics[scale=0.5]{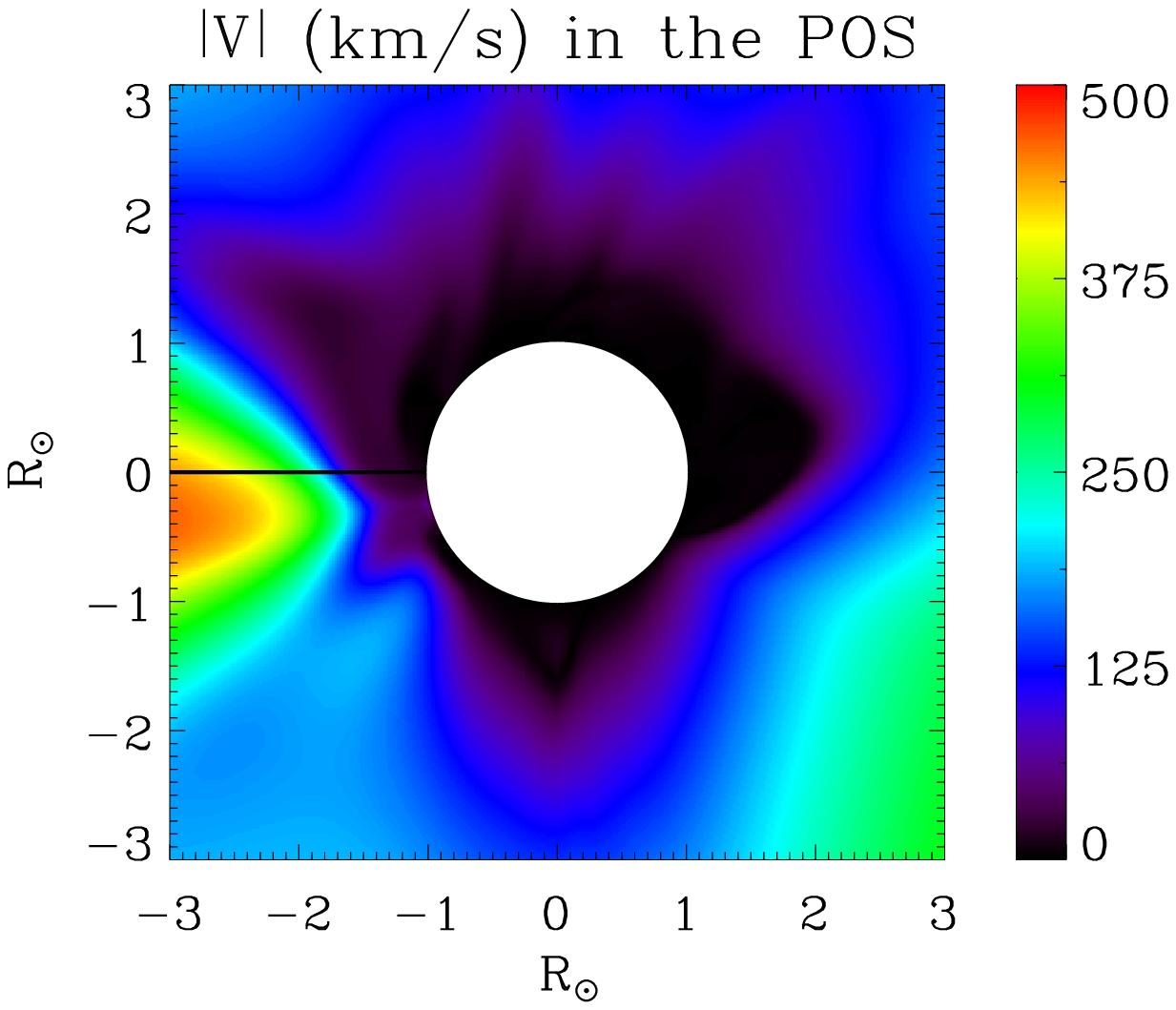}
	\caption{Temperature, proton number density, magnetic field strength, and
          modulus of the macroscopic velocity for ``the magnetic model'' CR2157
          in the POS. The solid line in the bottom right panel indicates the
          direction across which the variation of different quantities for this
          model are given in Figures \ref{fig:1dVandB}, \ref{fig:exposuretime}, and
          \ref{fig:number-density-photons}.}
	\label{fig:cr2157-param}
\end{figure*}

\begin{figure*}
	\includegraphics[scale=0.5]{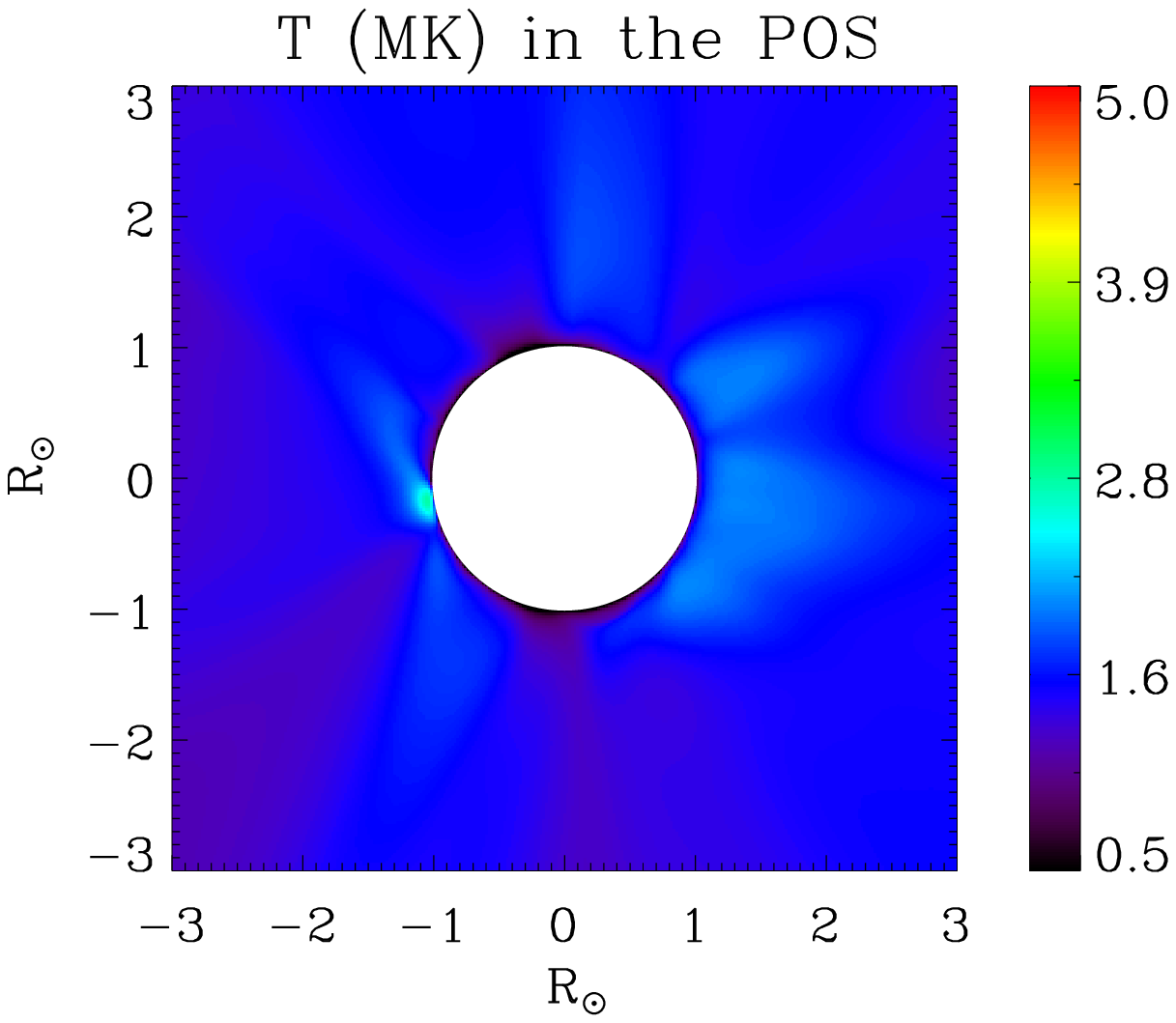}
	\includegraphics[scale=0.5]{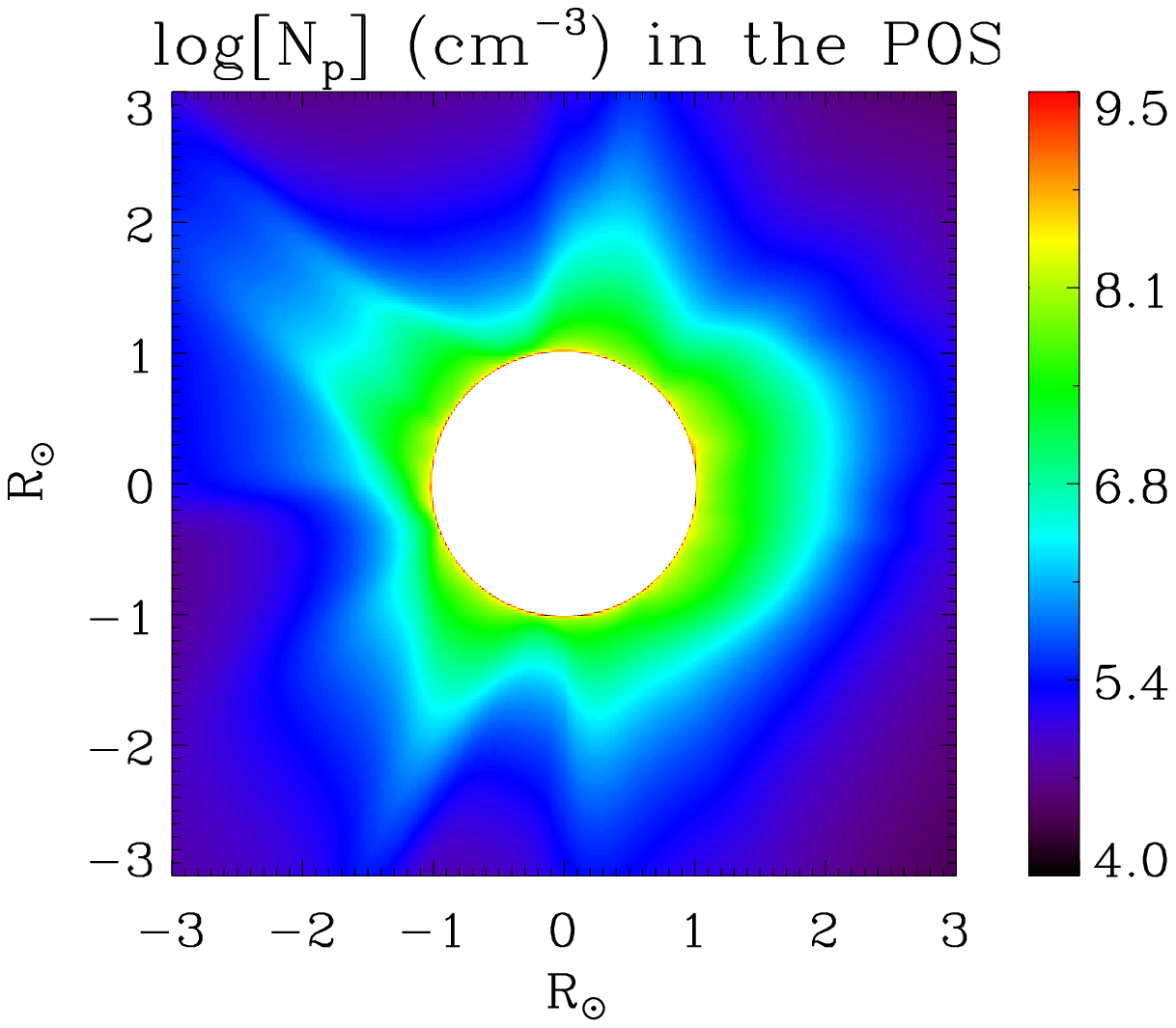}
	\includegraphics[scale=0.5]{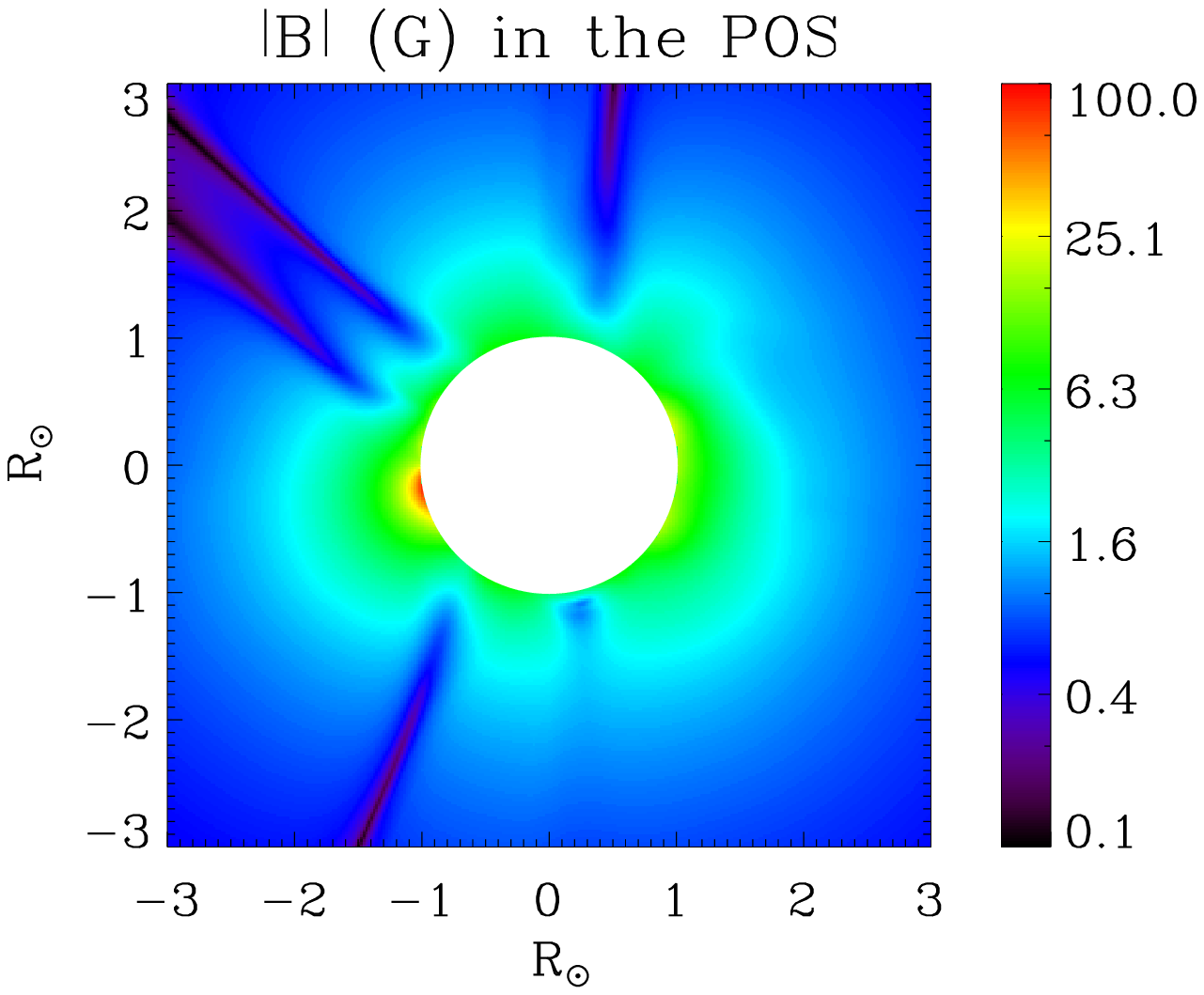}
	\includegraphics[scale=0.5]{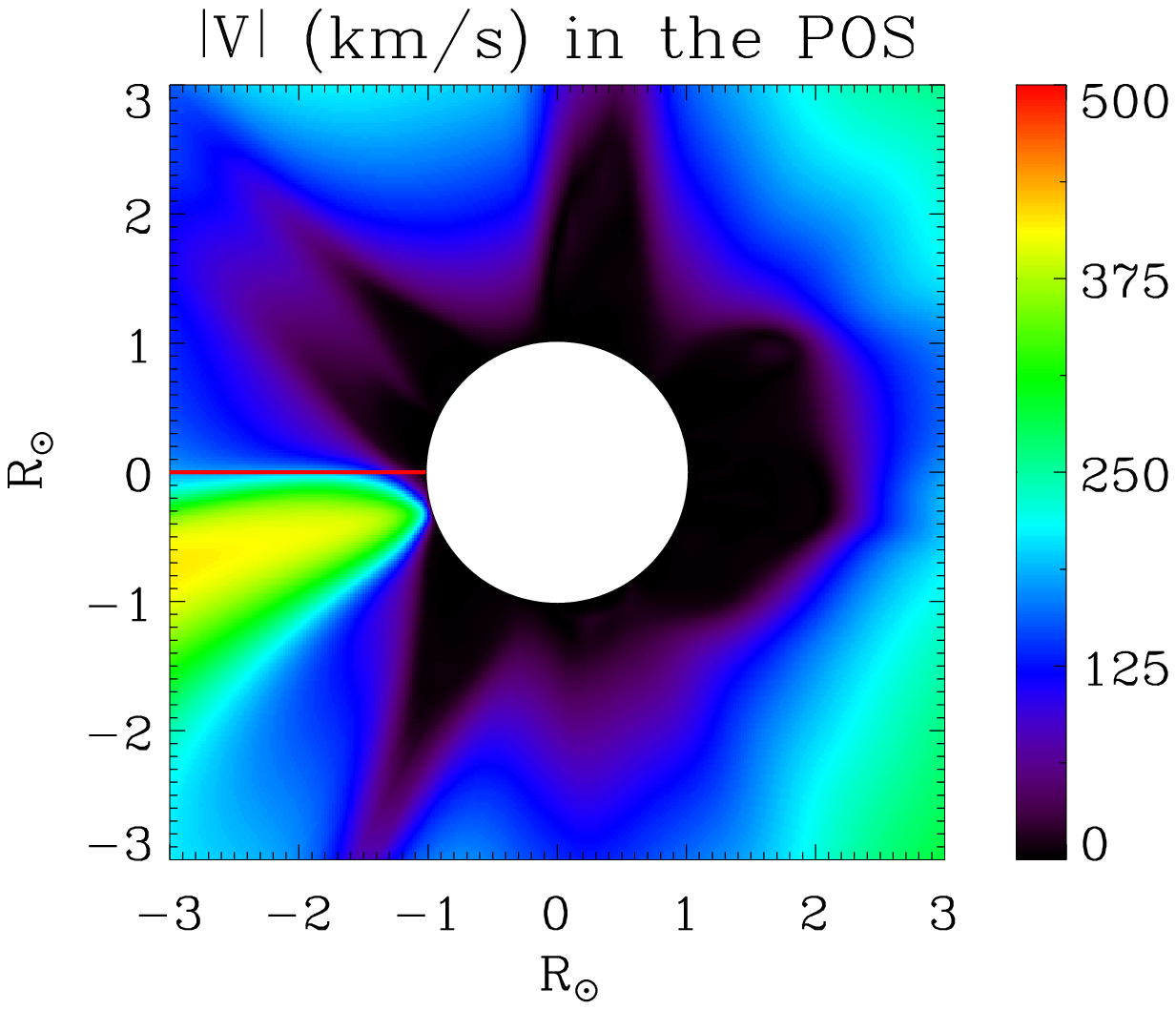}
	\caption{Temperature, proton number density, magnetic field strength, and
          modulus of the macroscopic velocity for ``the dynamic model'' CR2138
          in the POS. The solid line in the bottom right panel indicates the
          direction across which the variation of different quantities for this
          model are given in Figures \ref{fig:1dVandB}, \ref{fig:exposuretime}, and
          \ref{fig:number-density-photons}.}
	\label{fig:cr2138-param}
\end{figure*}

\begin{figure}
    \centering
    \includegraphics[scale=0.55]{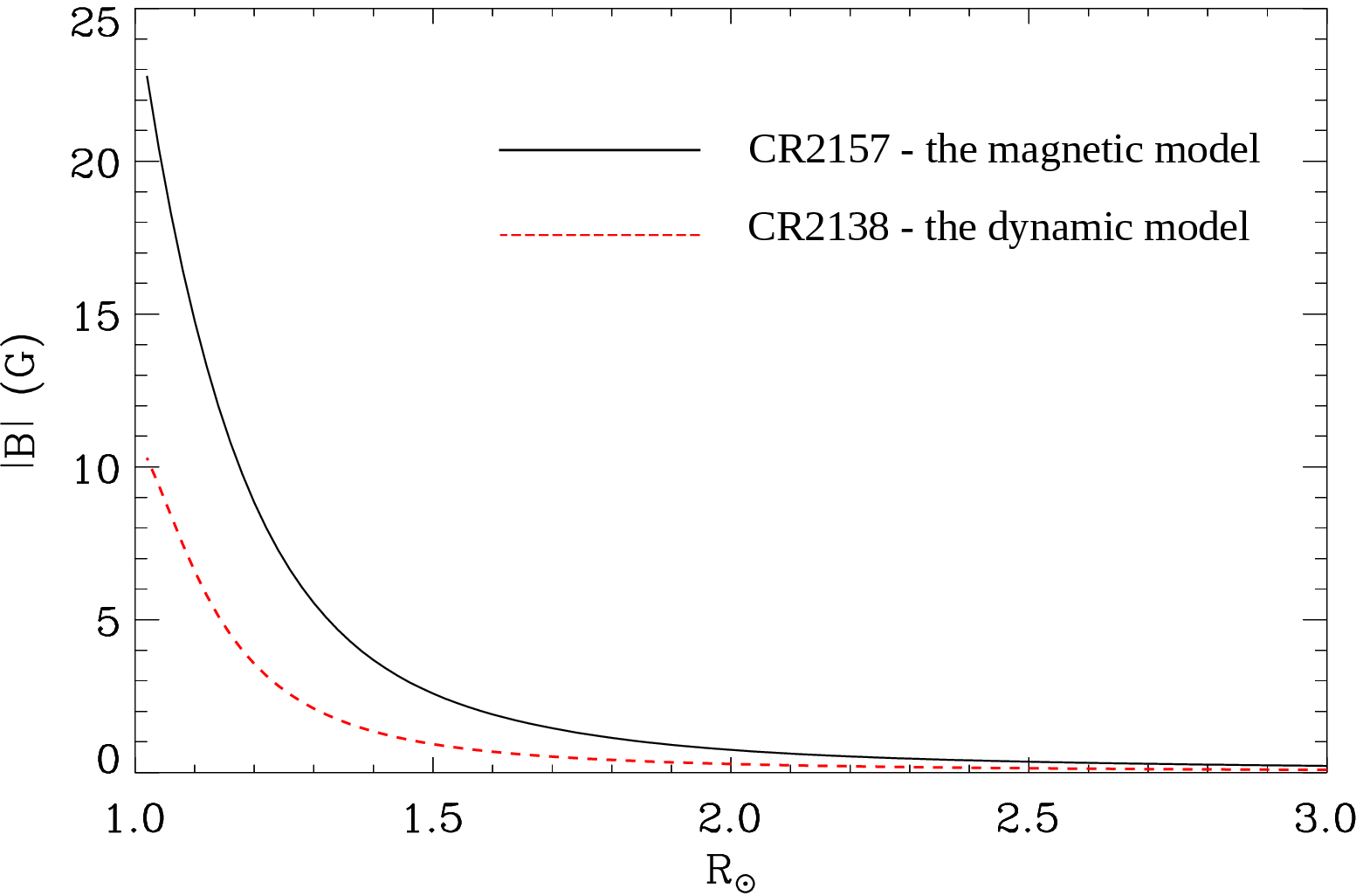}
    \includegraphics[scale=0.55]{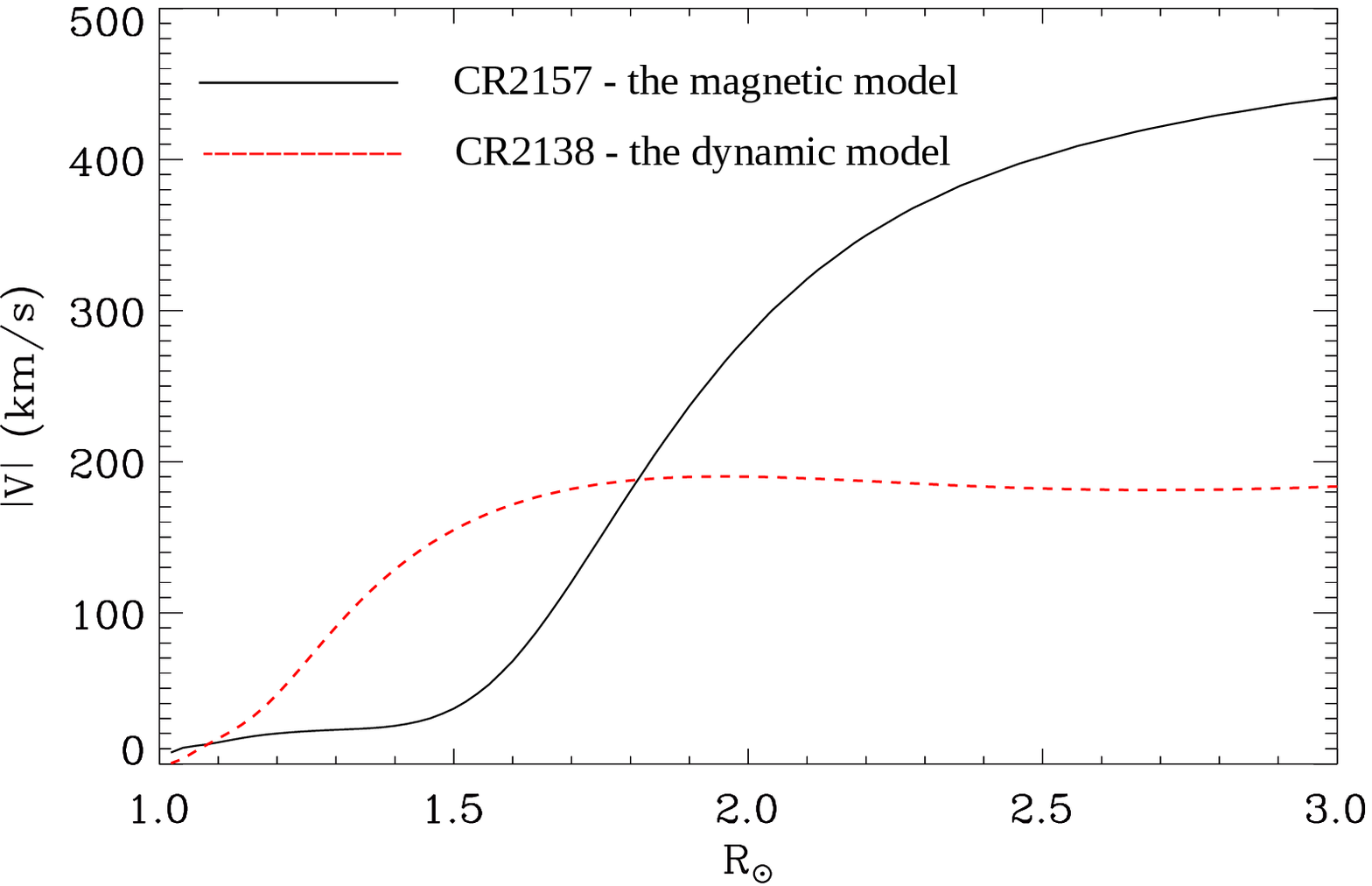}
    \caption{Variation of the model's magnetic field strength and modulus of the velocity along the radial direction  
    indicated in Figures \ref{fig:cr2157-param} and
    \ref{fig:cr2138-param}.}
    \label{fig:1dVandB}
\end{figure}

We use two of such coronal models: CR2138 (corresponding to 2013 June 30) 
and CR2157 (corresponding to 2014 November 14). Figures \ref{fig:cr2157-param} 
and \ref{fig:cr2138-param} visualize the temperature, the proton density, the magnetic field 
strength and the velocity of models CR2157 and CR2138, respectively. While the model CR2157  
shows a close to the limb ($h<$1.5R$_\odot$) magnetic activity larger than the model CR2138, it also  
presents weaker macroscopic velocities at such heights. Hereafter, we call the model CR2157 ``the magnetic model" (see Figure \ref{fig:cr2157-param}) and the model CR2138 ``the dynamic model" (see Figure \ref{fig:cr2138-param}). These two models 
are useful to illustrate the different sensitivities of the considered Ly-$\alpha$ lines to the 
coronal magnetic field and to the solar wind velocity. 
For a better comparison between the two coronal models, Figure \ref{fig:1dVandB} shows the 
variation of the magnetic field strength and modulus of the velocity 
along the radial directions indicated in the bottom right panel of 
Figures \ref{fig:cr2157-param} and \ref{fig:cr2138-param}.

The number densities of \ion{H}{1} and \ion{He}{2} are computed using the relation 
\begin{equation}
	N(\mathrm X^{+m}) = \frac{N(\mathrm X^{+m})}{N(\mathrm X)} 
	\frac{N(\mathrm X)}{N(\mathrm H)} N(\mathrm H),
\end{equation}  
where $N(\mathrm X)$, $N(\mathrm X^{+m})$, and $N(\mathrm H)$ 
are the number density of element X, the number density of element X in the $m$-th ionization
stage, and the hydrogen number density, respectively. 
The ionization fraction of the element X$^{+m}$, $N(\mathrm X^{+m})/N(\mathrm
X)$, is taken from the 
CHIANTI database - version 10 \citep{1997A&AS..125..149D,2021ApJ...909...38D}. The abundance ratio in the solar
corona, $N(\mathrm X)/N(\mathrm H)$, is as given in
\citet{2012ApJ...755...33S}. We assume that $N(\mathrm H)$ is the same as the proton number densities given in the 3D model being considered, i.e., fully ionized plasma.

In the following sections we show: 

\begin{itemize}

\item the total number of photons, $n_{\rm photons}$, emitted per unit area, per unit time, and per steradian, 

\item the total linear polarization, $\textnormal P=\frac{\sqrt{\textnormal Q^2+\textnormal U^2}}{\textnormal I}$,

\item the relative polarization, ${\rm P}_{\rm r}=\frac{\textnormal P_0-\textnormal P}{\textnormal P_0}$ (with  P$_0$ the degree of polarization when there are no symmetry breaking effects), 

\item the rotation of the polarization plane, $\textnormal R = \frac{1}{2} \textnormal {tan}^{-1}\big(\frac{\textnormal U}{\textnormal Q}\big)$, with respect to the local solar limb. 

\end{itemize}
The linear polarization maps shown below 
correspond to a LOS integration of 6R$_\odot$ centered around the POS. 

\subsubsection{Polarization maps of ``the magnetic model'' CR2157}

Some physical quantities of this coronal model are shown in Figure \ref{fig:cr2157-param}, for positions in 
the POS up to approximately two solar radii above the model's visible disk. 
The application of our computer code for calculating the polarization of the spectral line radiation scattered by the 
solar corona allows us to perform several numerical experiments, useful for understanding 
the diagnostic potential of the lines under consideration. We recall that this requires to calculate the comoving 
frame radiation field tensor of the spectral line at each spatial point within the corona of the model, to 
solve the statistical equilibrium equations to determine the emission coefficient in the Stokes parameters, and to 
calculate the emergent Stokes signals after integrating along each LOS and over the line's frequency interval. 

\begin{figure*}
\includegraphics[scale=0.5]{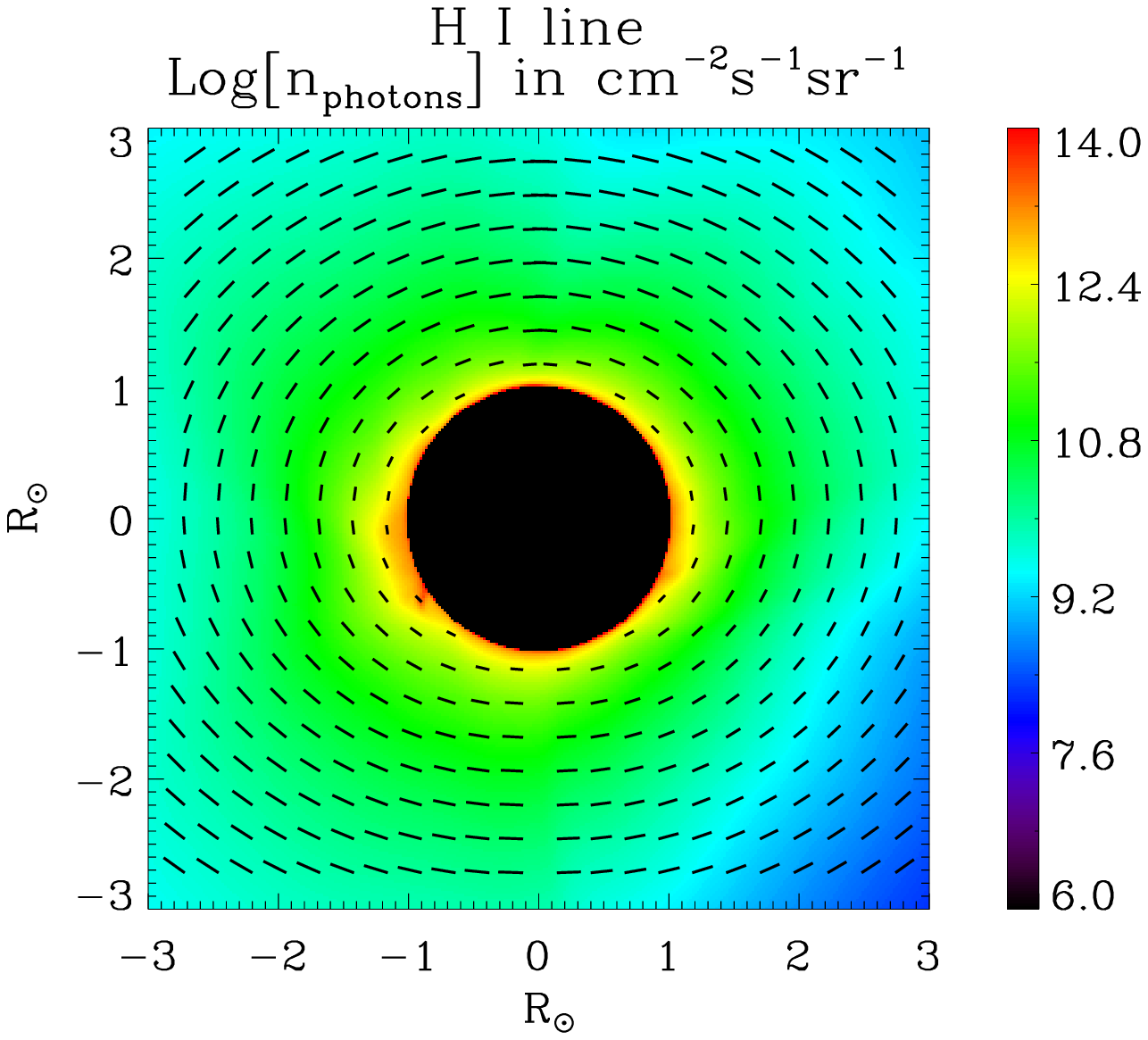}
\includegraphics[scale=0.5]{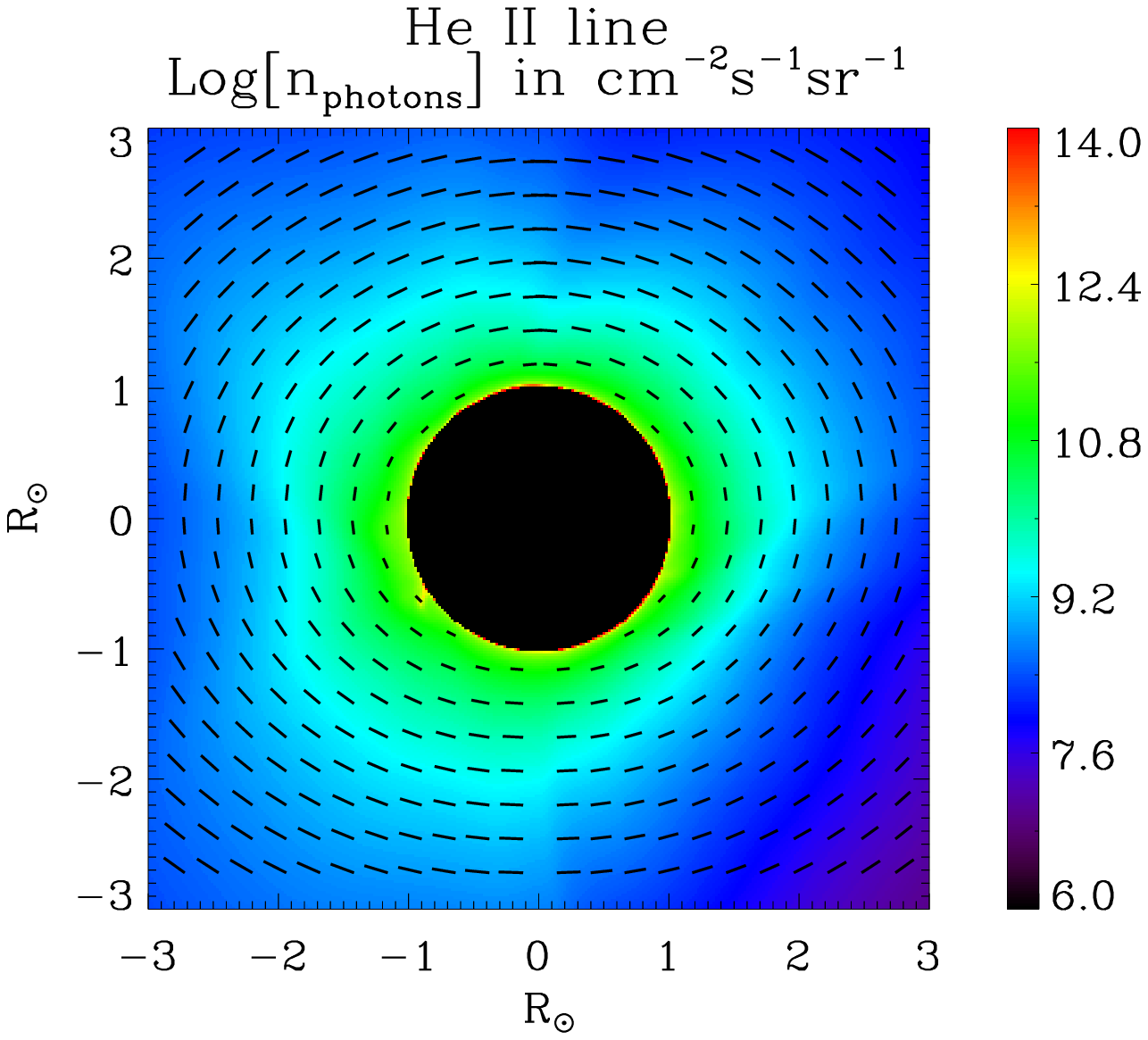}
\caption{The intensity of the Ly-$\alpha$ lines of
H {\sc i} (left panel) and He {\sc ii} (right panel) computed in ``the magnetic model'' CR2157 
without its magnetic field 
and macroscopic velocity, including the LOS integration. The black short lines 
indicate the direction of the linear polarization and their length the amplitude of the polarization signals.}
\label{fig:cr2157-intensity}
\end{figure*}

\begin{figure*}
\includegraphics[scale=0.5]{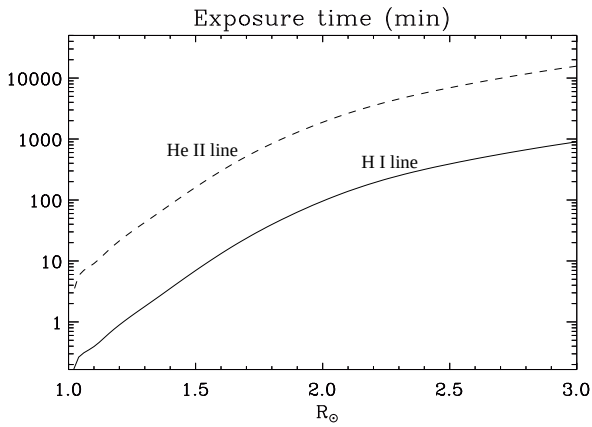}
\caption{Required exposure time for the Ly-$\alpha$ lines of
H {\sc i} and He {\sc ii} along the radial direction computed for a set of parameters specified in the text.}
\label{fig:exposuretime}
\end{figure*}

We start by considering the nonmagnetized and static case. To this end, we carried out 
the calculations after forcing to zero the model's magnetic field 
and macroscopic velocity. Figure \ref{fig:cr2157-intensity} shows the results for the number of photons 
in the Ly-$\alpha$ lines of \ion{H}{1} (left panel) and \ion{He}{2} (right panel), per unit area, per unit time and 
per solid angle unit. In these panels the overplotted short black lines indicate the direction of the 
linear polarization signals and their length the polarization amplitude. 
The number of photons in the 
line radiation scattered by the model's corona is larger in the 
hydrogen Ly-$\alpha$ line despite the fact that the number density of residual neutral hydrogen atoms 
is slightly less than the number density of \ion{He}{2} atoms (see the top panel of Figure 
\ref{fig:number-density-photons}). This is because of the larger incoming radiation for the Ly-$\alpha$ 
line of \ion{H}{1} as compared to that of the Ly-$\alpha$ 
line of \ion{He}{2}.
As seen in the right panel of Figure \ref{fig:cr2157-intensity} and the bottom panel of 
Figure \ref{fig:number-density-photons}, the number of photons in the 
Ly-$\alpha$ line of \ion{He}{2} are very low after 0.5R$_\odot$ above 
the model's visible disk; hence, detecting the polarization of this line at those heights would require 
prohibitively long integration times and/or telescope apertures. To substantiate this, in Figure 
\ref{fig:exposuretime} we compare the 
exposure times, for both the intensities in Figure \ref{fig:cr2157-intensity}, needed to measure 10$^7$ photons with an instrument of 500~cm$^2$ collection area, 20 arcsec spatial 
sampling and an instrument efficiency of 0.01. 
For this reason, the remaining figures show the results up to 0.5R$_\odot$ above the model's visible disk.

\begin{figure*}
	\includegraphics[scale=0.5]{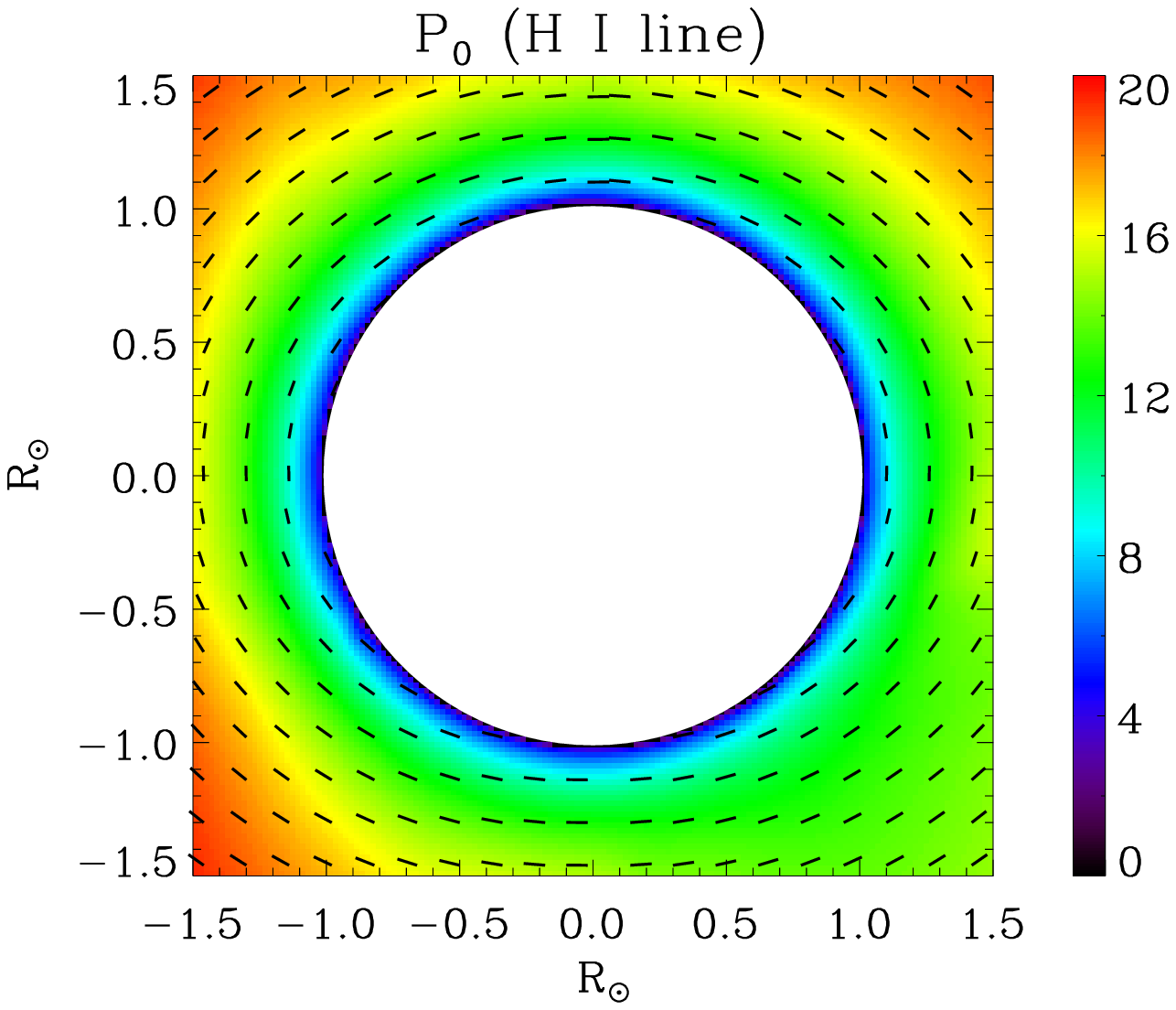}
	\includegraphics[scale=0.5]{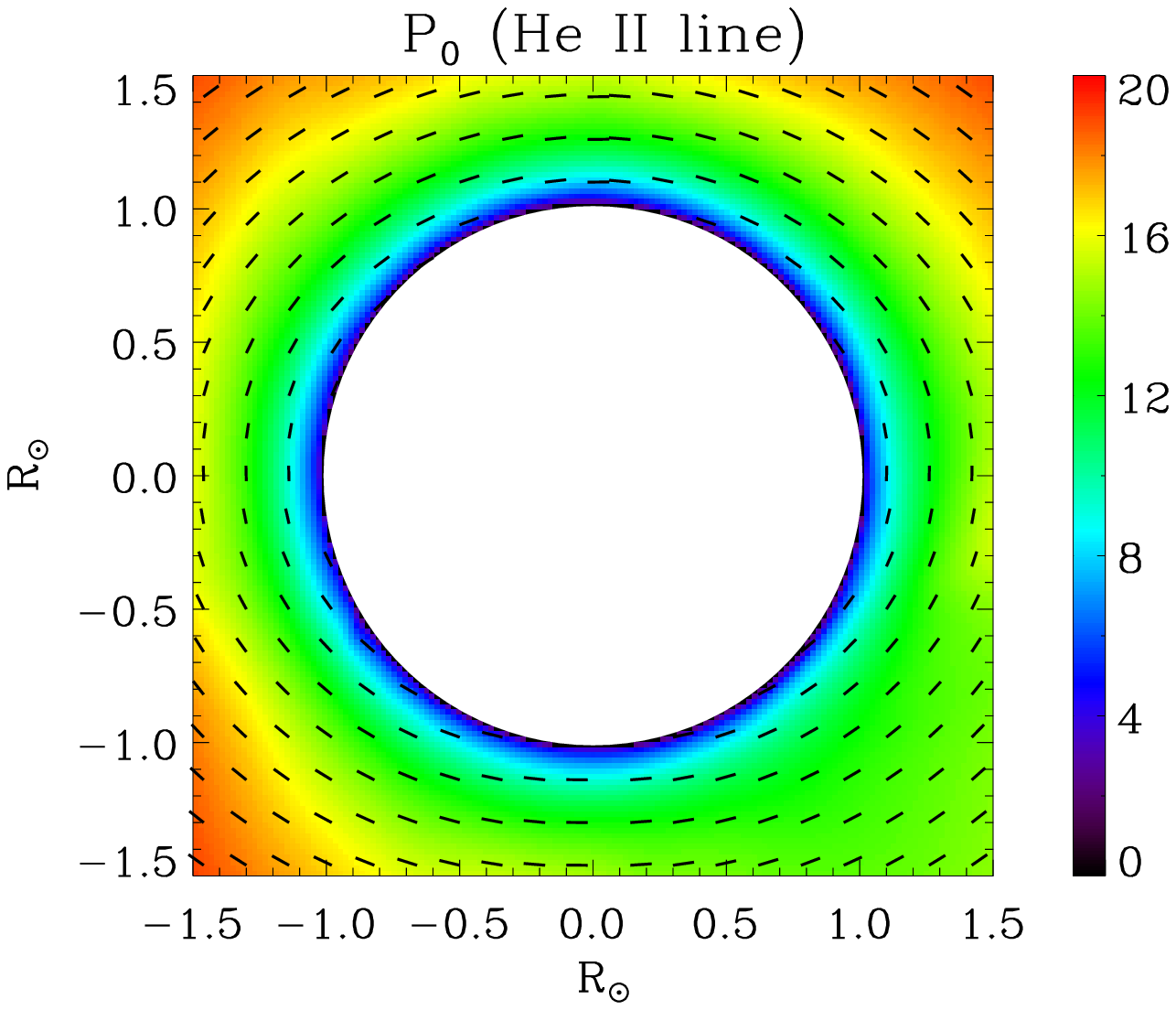}
\caption{The fractional total linear polarization of the Ly-$\alpha$ lines of
H {\sc i} (left panel) and He {\sc ii} (right panel) computed in ``the magnetic model'' 
CR2157 without its magnetic field 
and macroscopic velocity, including the LOS integration. Like in 
Figure~\ref{fig:cr2157-intensity}, the black short lines 
indicate the direction of the linear polarization and their length the amplitude of the polarization signals.}
\label{fig:cr2157-polarization-SN}
\end{figure*}

As expected, in the absence of any symmetry breaking, 
the linear polarization of the scattered radiation is always perpendicular to the solar radius 
vector through the observed point. Moreover, the amplitude of the linear polarization is the same 
in both Ly-$\alpha$ lines, because their levels have the same angular momentum values and, therefore, 
the same polarizability. This can also be seen in Figure \ref{fig:cr2157-polarization-SN}, which shows that the 
total fractional linear polarization signals for the non-magnetic and static 
case under consideration increase with 
height in the solar coronal model, reaching values of about 20\% at heights 
$h\,{\approx}\,0.5{\rm R}_\odot$ above the solar surface.

\begin{figure*}
	\includegraphics[scale=0.5]{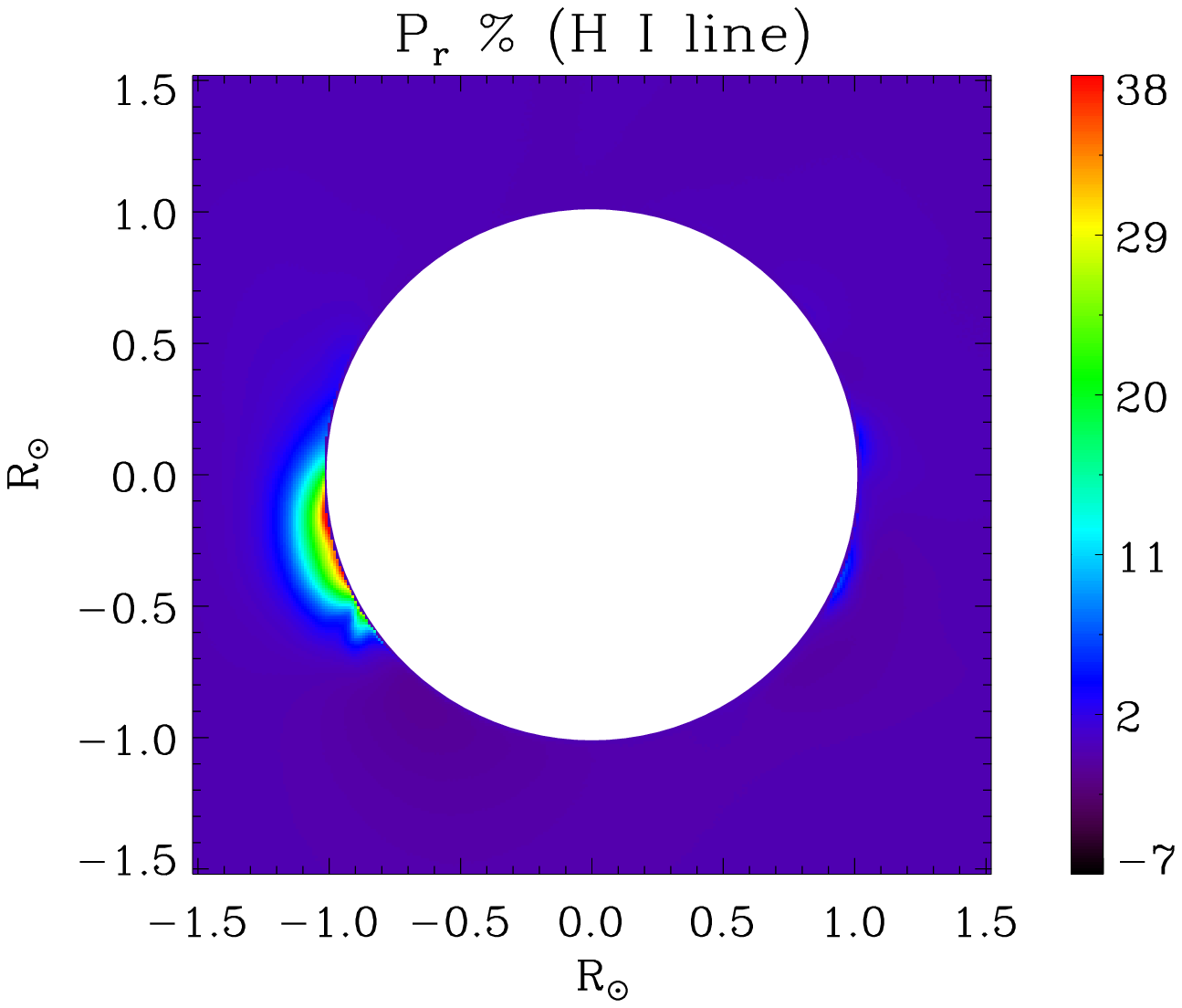}
	\includegraphics[scale=0.5]{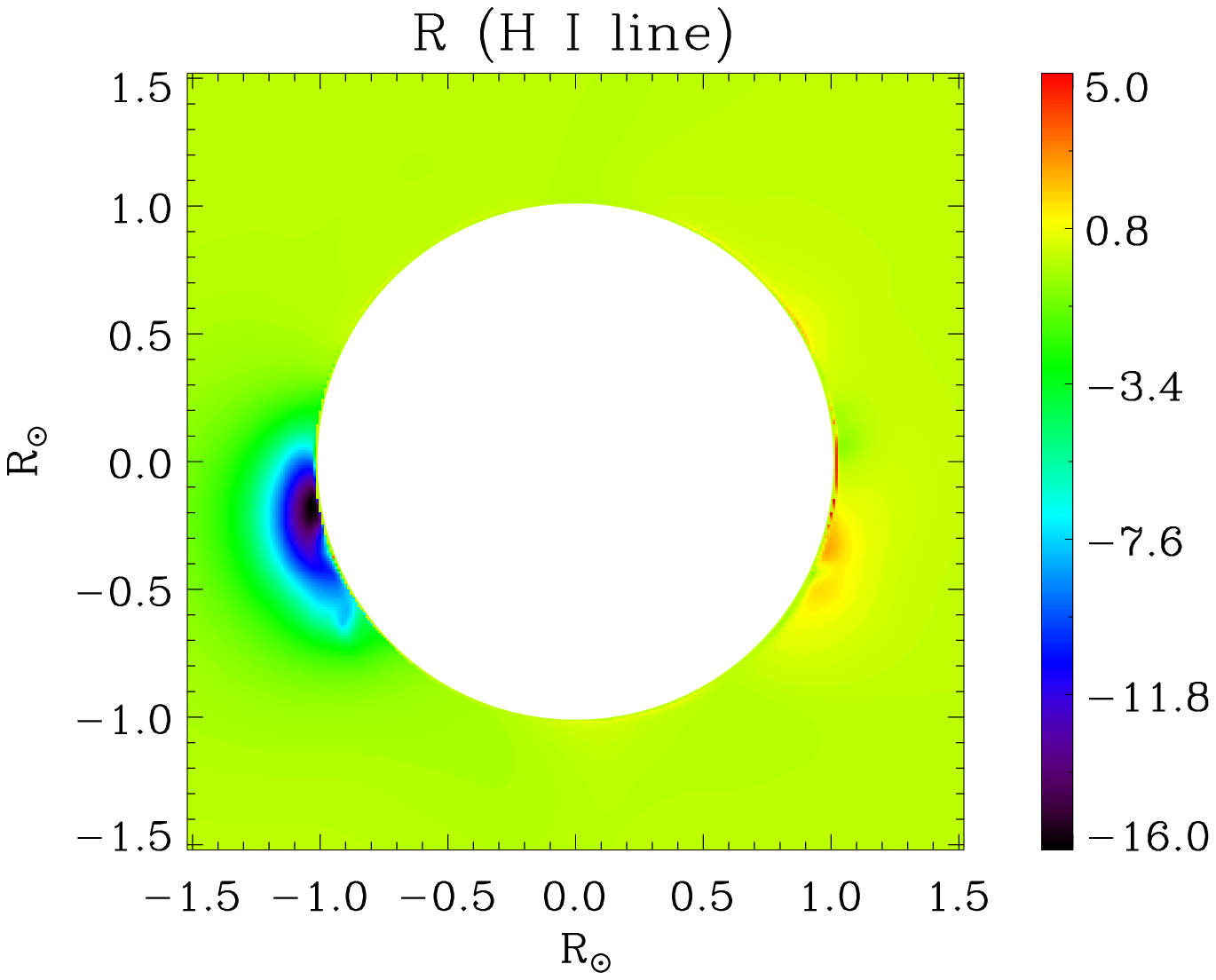}
	\includegraphics[scale=0.5]{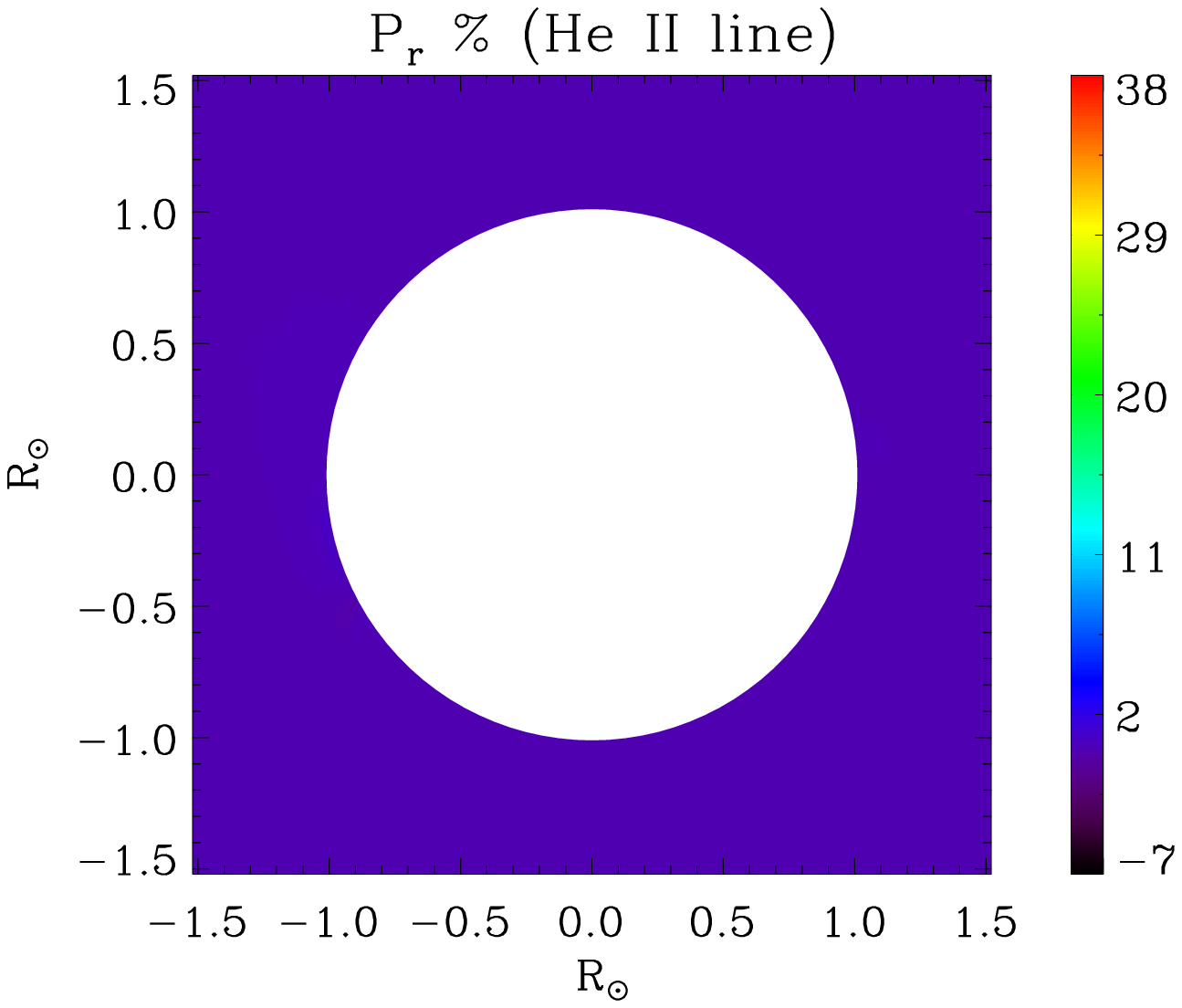}
	\includegraphics[scale=0.5]{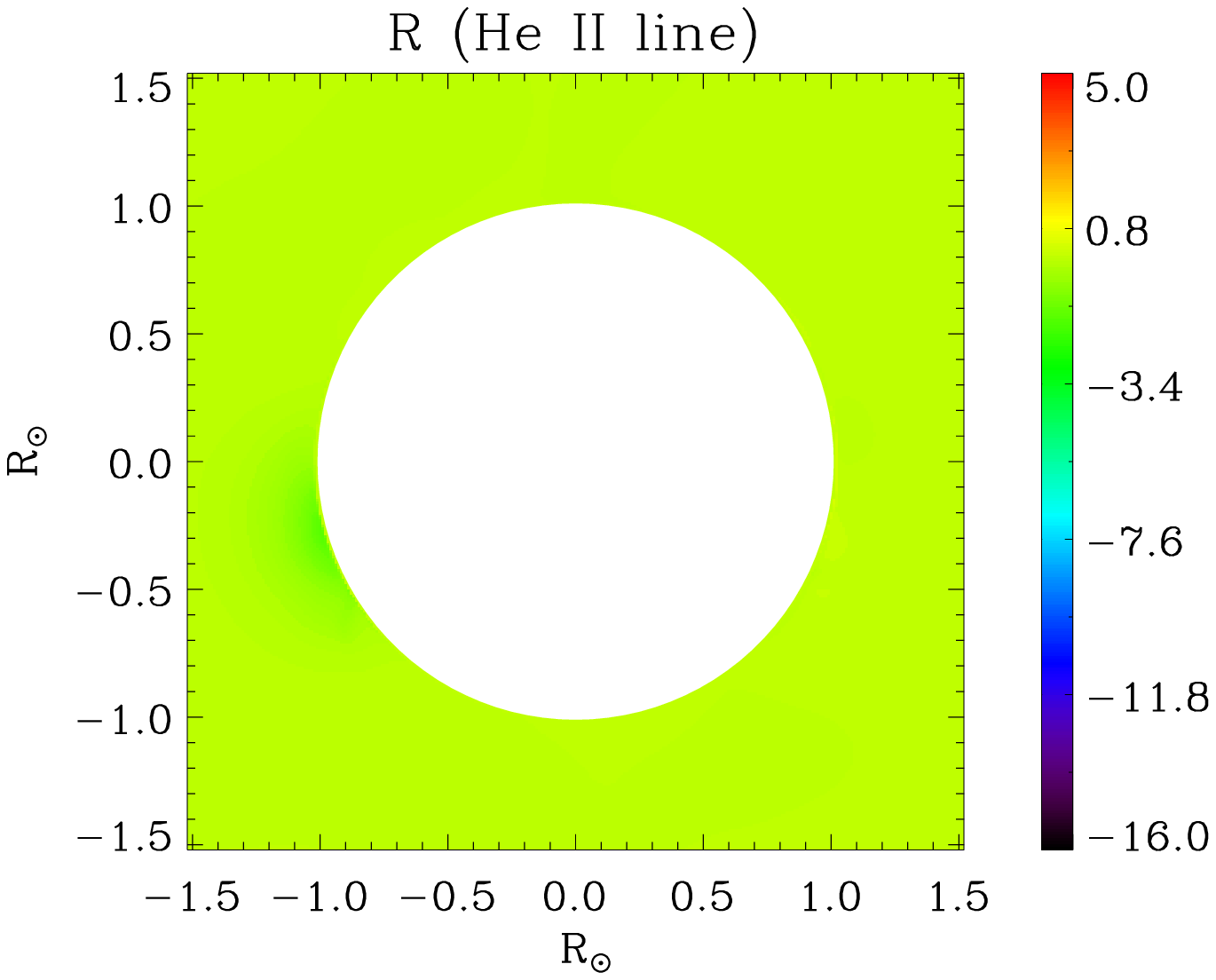}
	\caption{The relative polarization $P_{\rm r}$ and the rotation angle ${\rm R}$ of the polarization plane
	of the Ly-$\alpha$ lines of H {\sc i} (top panels) and the He {\sc ii} (bottom panels) calculated in 
	``the magnetic model'' CR2157 taking into account the Hanle effect but ignoring the model's macroscopic velocity.}
	\label{fig:cr2157-reldepolrot-SM}
\end{figure*}

In Figure \ref{fig:cr2157-reldepolrot-SM} we show what happens when we take into account only the Hanle effect produced by the model's magnetic 
field (i.e., we force to zero the model's macroscopic velocity, thus 
assuming that there is no solar wind). The figure shows the relative polarization   
${\rm P}_{\rm r}$ (left panels) and the rotation of the polarization plane $\textnormal R$ for the Ly-$\alpha$ lines 
of \ion{H}{1} (top panels) and \ion{He}{2} (bottom panels). Obviously, in the static 
case being considered ${\rm P}_{\rm r}=\textnormal R=0$, where the Hanle effect does not operate.
As mentioned above, the critical magnetic field for the onset of the Hanle effect is $B_{\rm H}=53$~G for
the hydrogen Ly-$\alpha$ line and $B_{\rm H}=850$~G for the Ly-$\alpha$ line of \ion{He}{2}, and the 
typical sensitivity to the Hanle effect occurs for magnetic strengths between 0.2$B_{\rm H}$ and 5$B_{\rm H}$. 
Given that the magnetic field strengths in ``the magnetic model'' CR2157 of the solar corona 
are weaker than 100~G, it is logical to find in Figure \ref{fig:cr2157-reldepolrot-SM} that 
only the hydrogen Ly-$\alpha$ line shows a significant Hanle effect. In particular, in the near to the limb regions 
where the model's magnetic field is more intense and/or longitudinal, 
the Hanle effect in the Ly-$\alpha$ line of \ion{H}{1} produces 
a sizable depolarization and rotation of the polarization plane. 

\begin{figure*}[!]
	\includegraphics[scale=0.5]{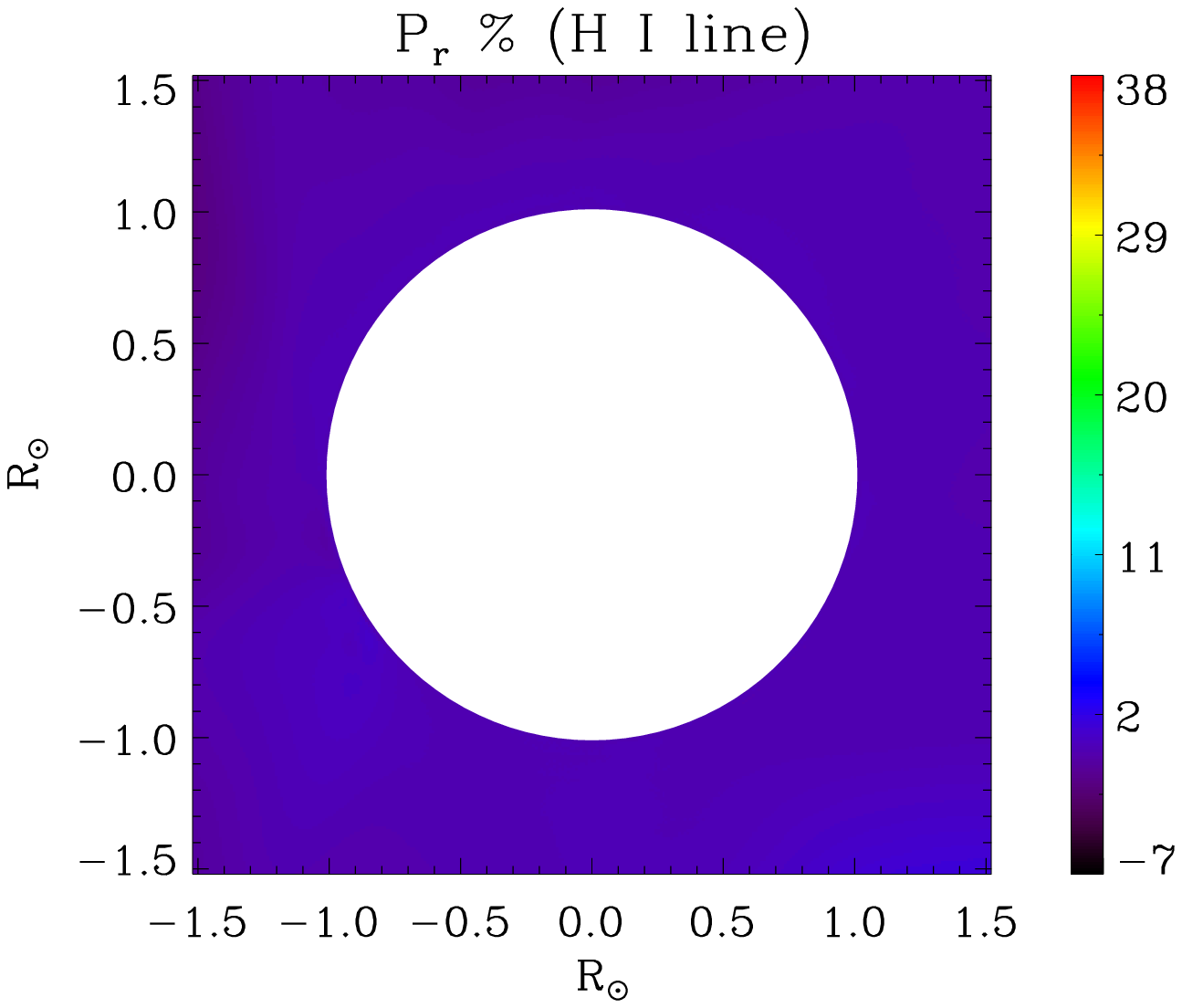}	
	\includegraphics[scale=0.5]{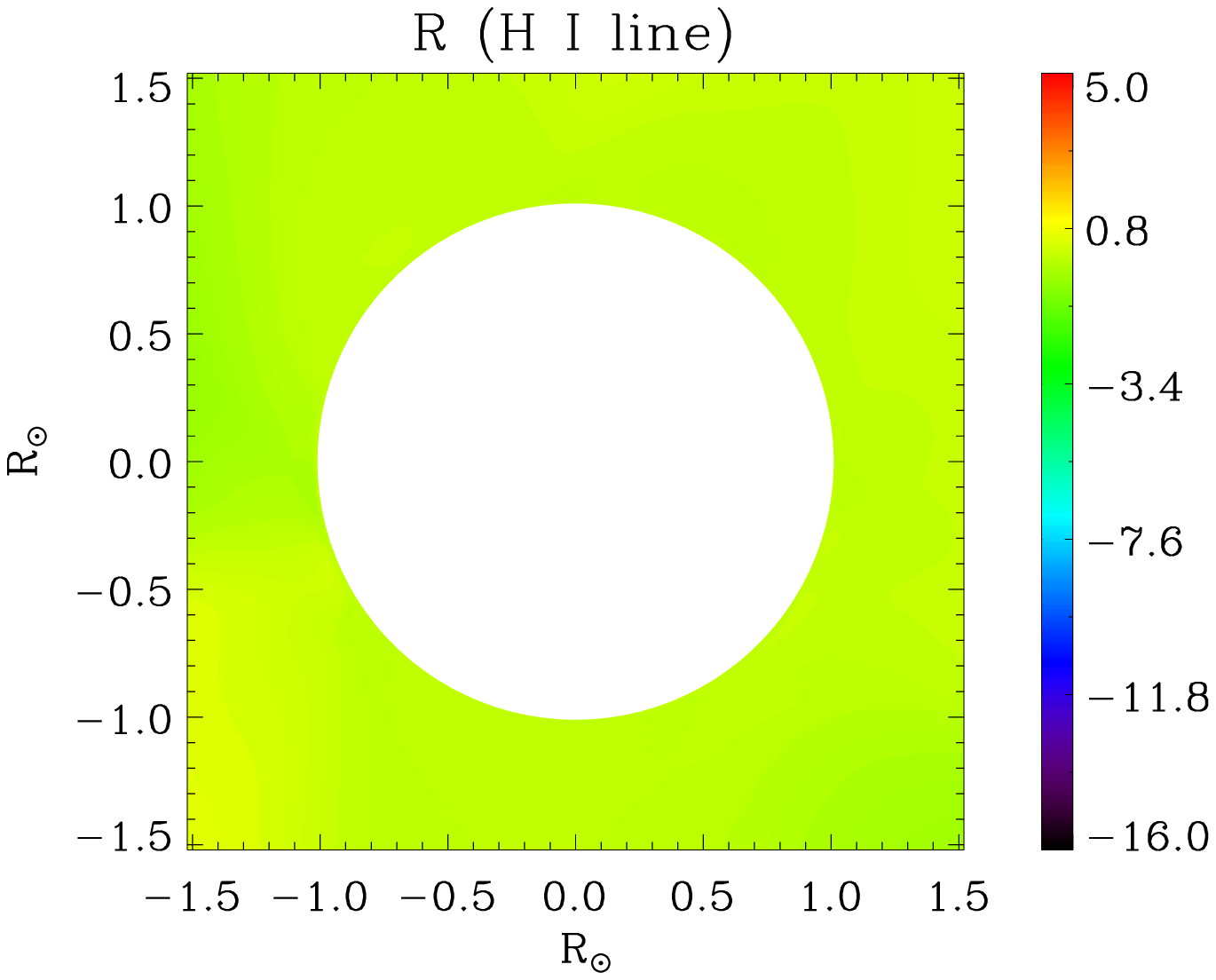}
	\includegraphics[scale=0.5]{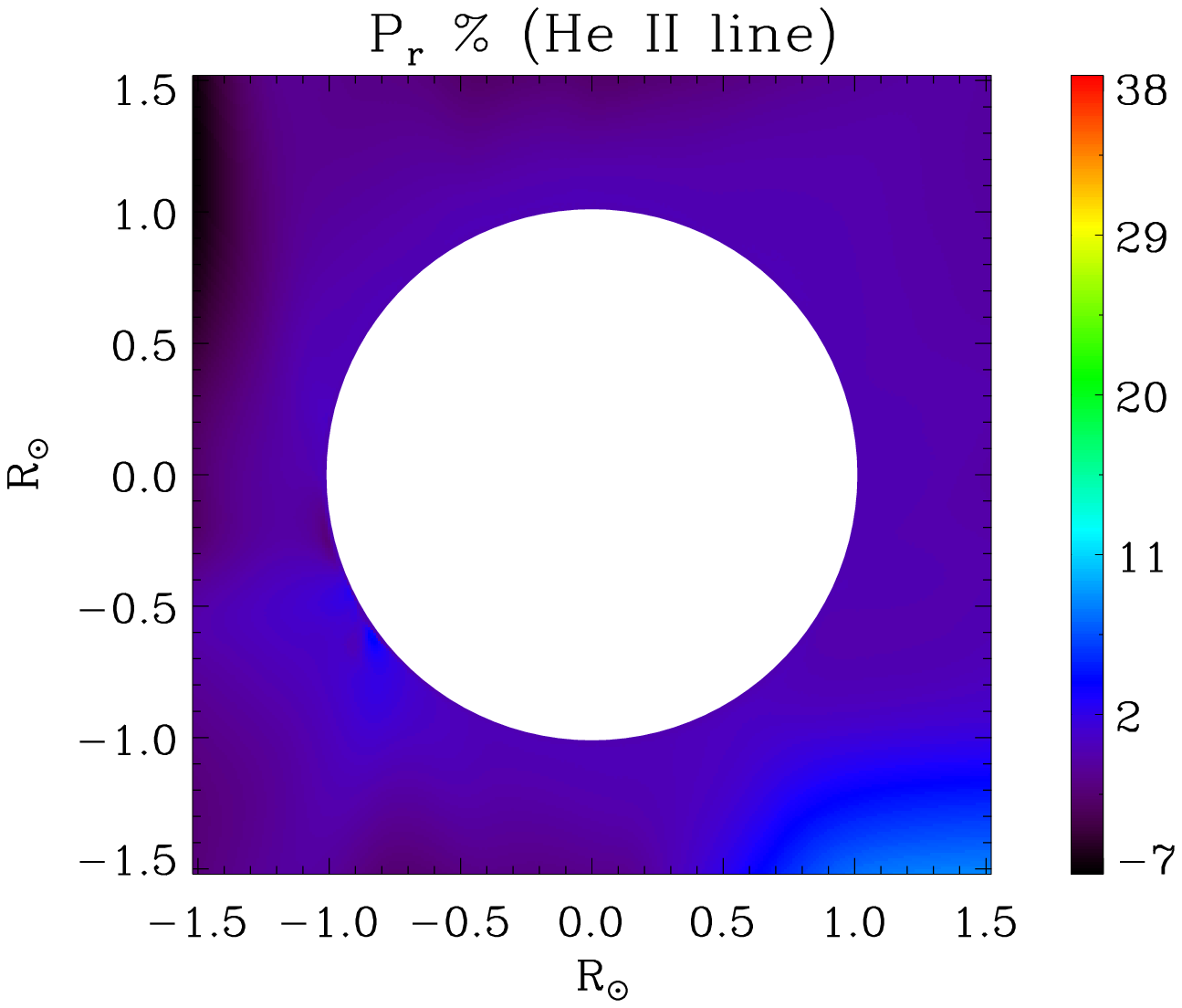}	
	\includegraphics[scale=0.5]{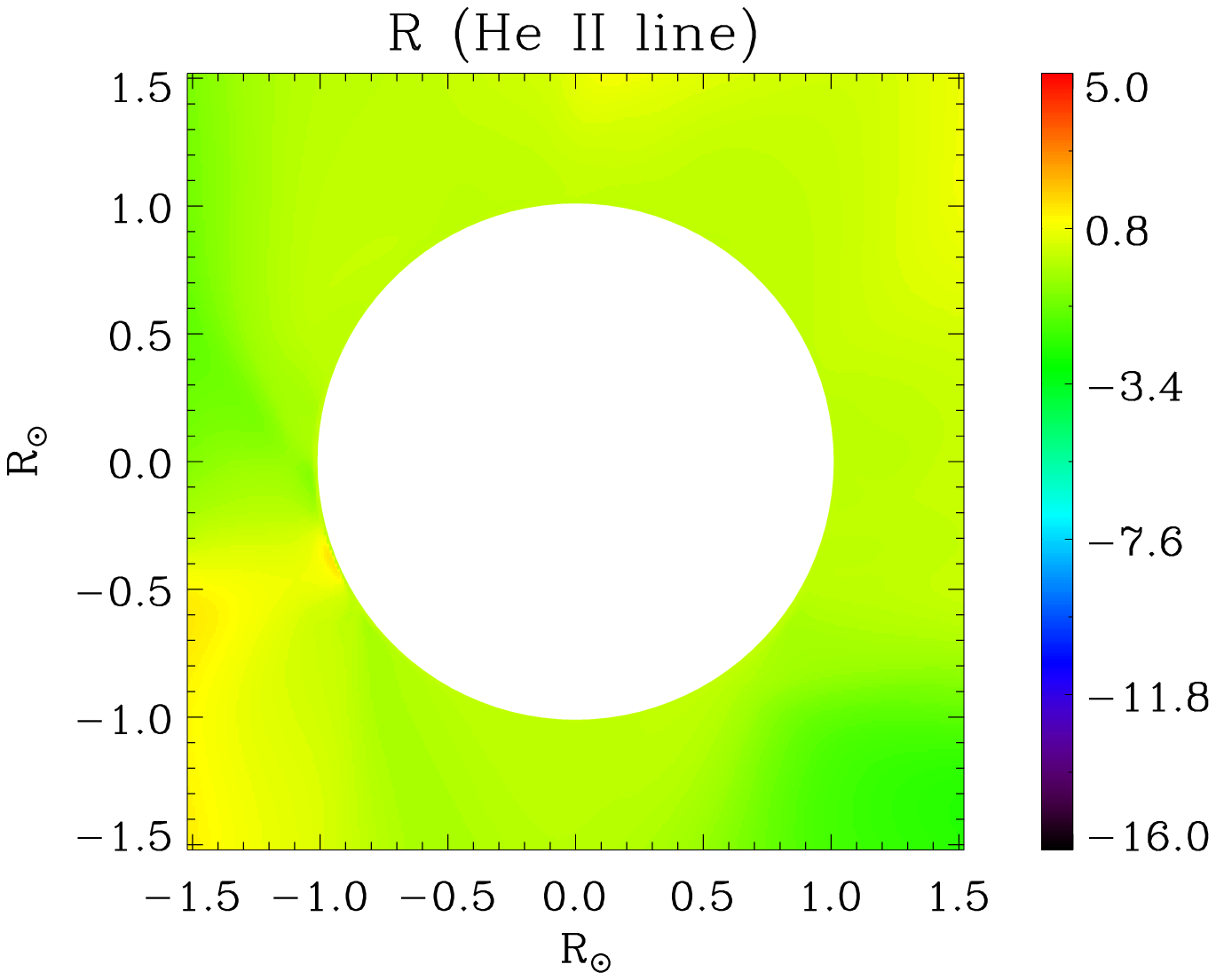}
	\caption{The relative polarization $P_{\rm r}$ and the rotation angle ${\rm R}$ of the polarization plane
	of the Ly-$\alpha$ lines of H {\sc i} (top panels) and the He {\sc ii} (bottom panels) calculated in  
	``the magnetic model'' CR2157 ignoring the Hanle effect but taking into account the impact of the 
	model's macroscopic velocity.}
	\label{fig:cr2157-reldepolrot-SV}
\end{figure*}

\begin{figure*}
    \centering
    \includegraphics[scale=0.5]{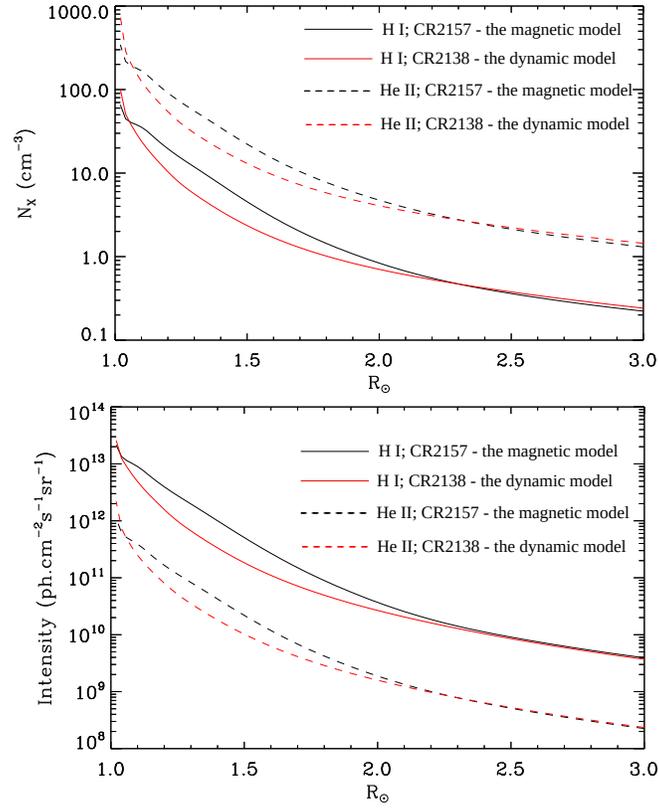}
    \caption{Variation of the model's number density of H {\sc i} and He {\sc ii} and of the 
    ensuing Lyman-$\alpha$ line intensities 
    along the radial direction indicated in Figures \ref{fig:cr2157-param} and \ref{fig:cr2138-param}.
    Both these quantities are obtained after the LOS integration.}
    \label{fig:number-density-photons}
\end{figure*}

Correspondingly, Figure \ref{fig:cr2157-reldepolrot-SV} isolates the impact of the Doppler effect produced 
by the solar wind of the model (i.e., we force to zero the magnetic field).  
Because the \ion{He}{2} Ly-$\alpha$ line coming from the underlying atmosphere is much narrower than the \ion{H}{1}
Ly-$\alpha$ line, the impact on the scattering polarization is much more significant for the \ion{He}{2} Ly-$\alpha$
line. Nevertheless, since below 1.5R$_\odot$ the 
model's macroscopic velocity is rather small (see the bottom right panels of Figures \ref{fig:cr2157-param} and \ref{fig:1dVandB}), 
the impact of the model's solar wind on the linear polarization of the \ion{He}{2} line is correspondingly 
small. In coronal regions similar to those in this model, the \ion{He}{2} 
Ly-$\alpha$ line can be considered to be a useful reference line for facilitating the detection of the 
fingerprints of the Hanle effect in the hydrogen Ly-$\alpha$ line.
                     
\subsubsection{Polarization maps of ``the dynamic model'' CR2138}

\begin{figure*}
	\includegraphics[scale=0.5]{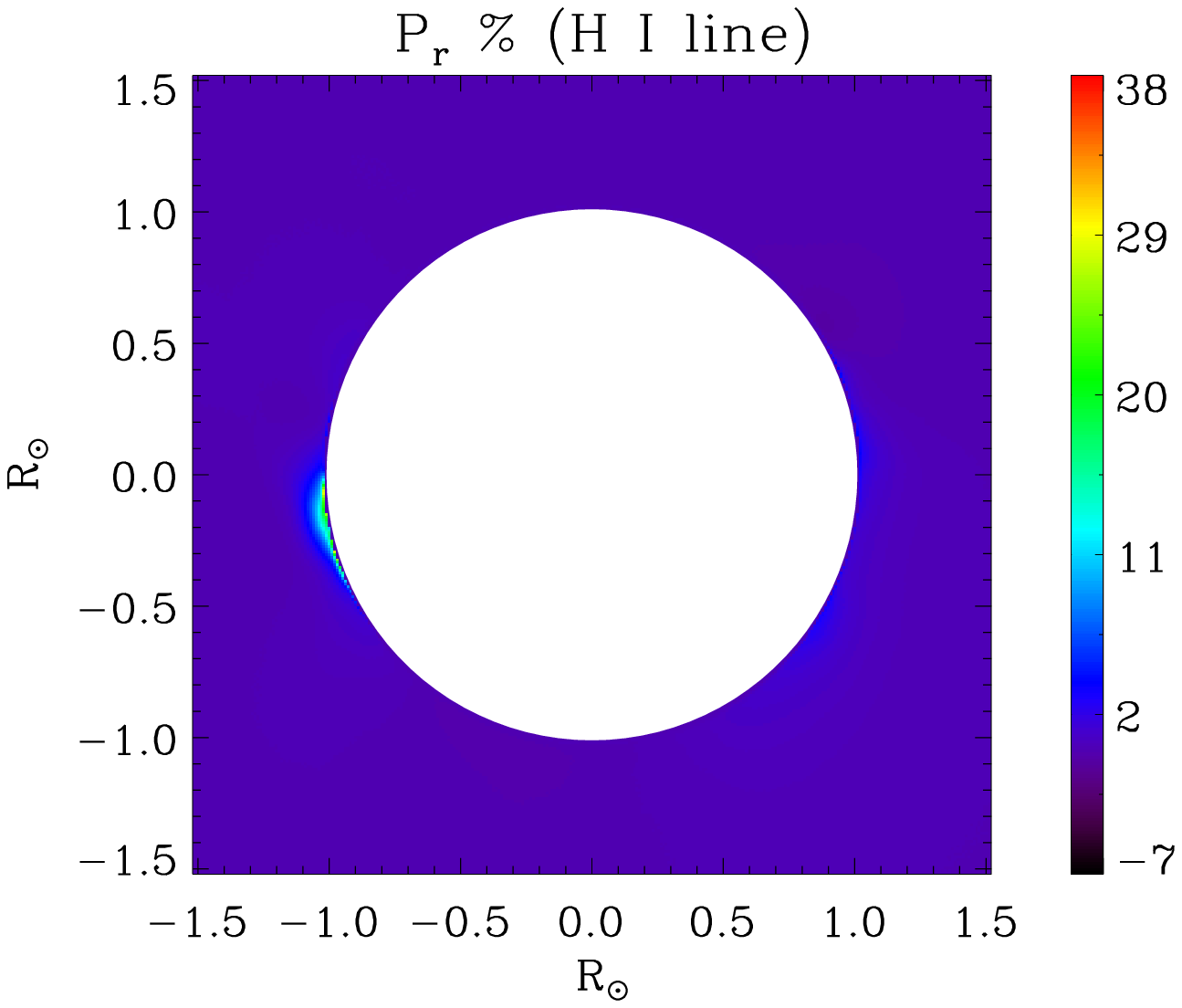}
	\includegraphics[scale=0.5]{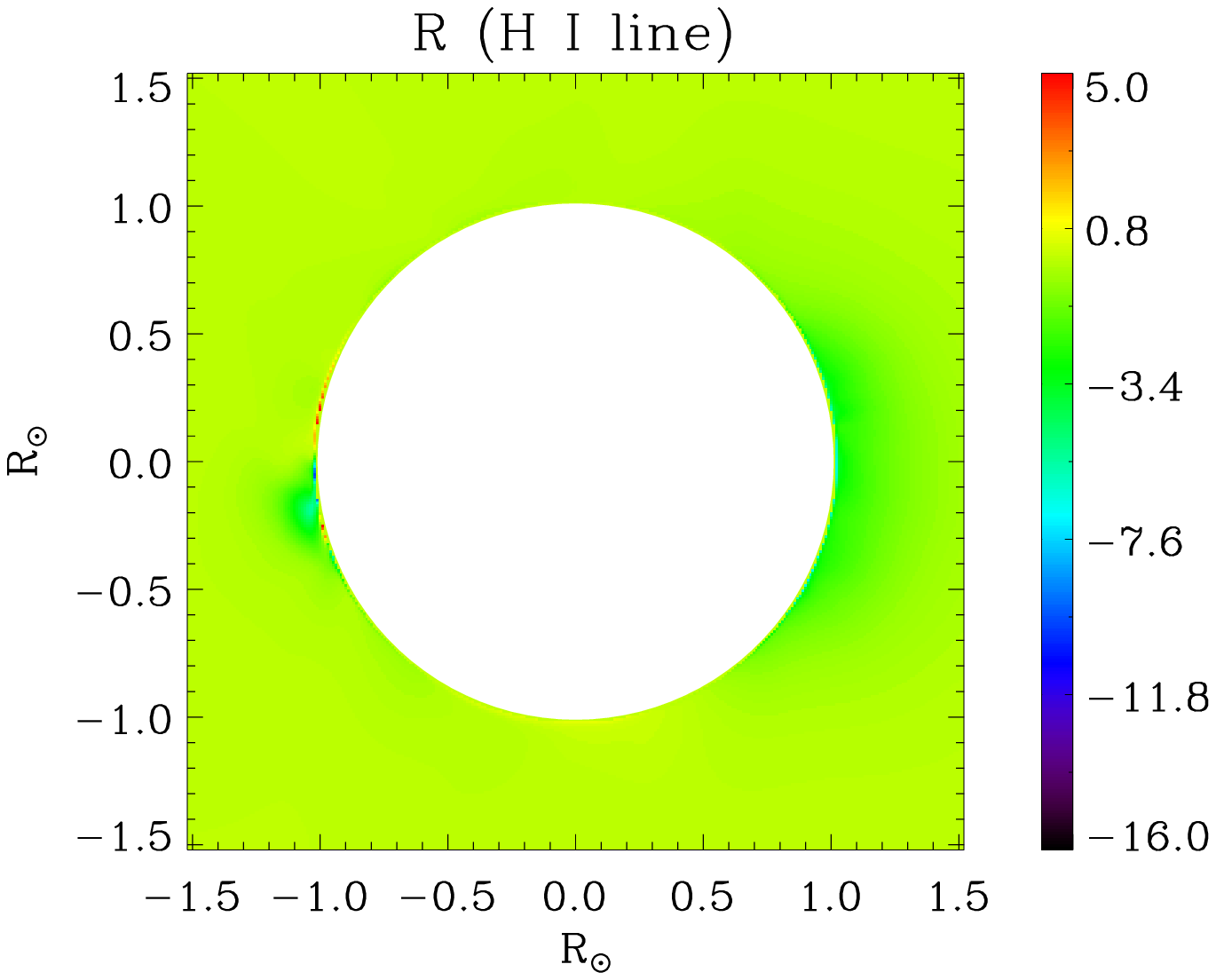}
	\includegraphics[scale=0.5]{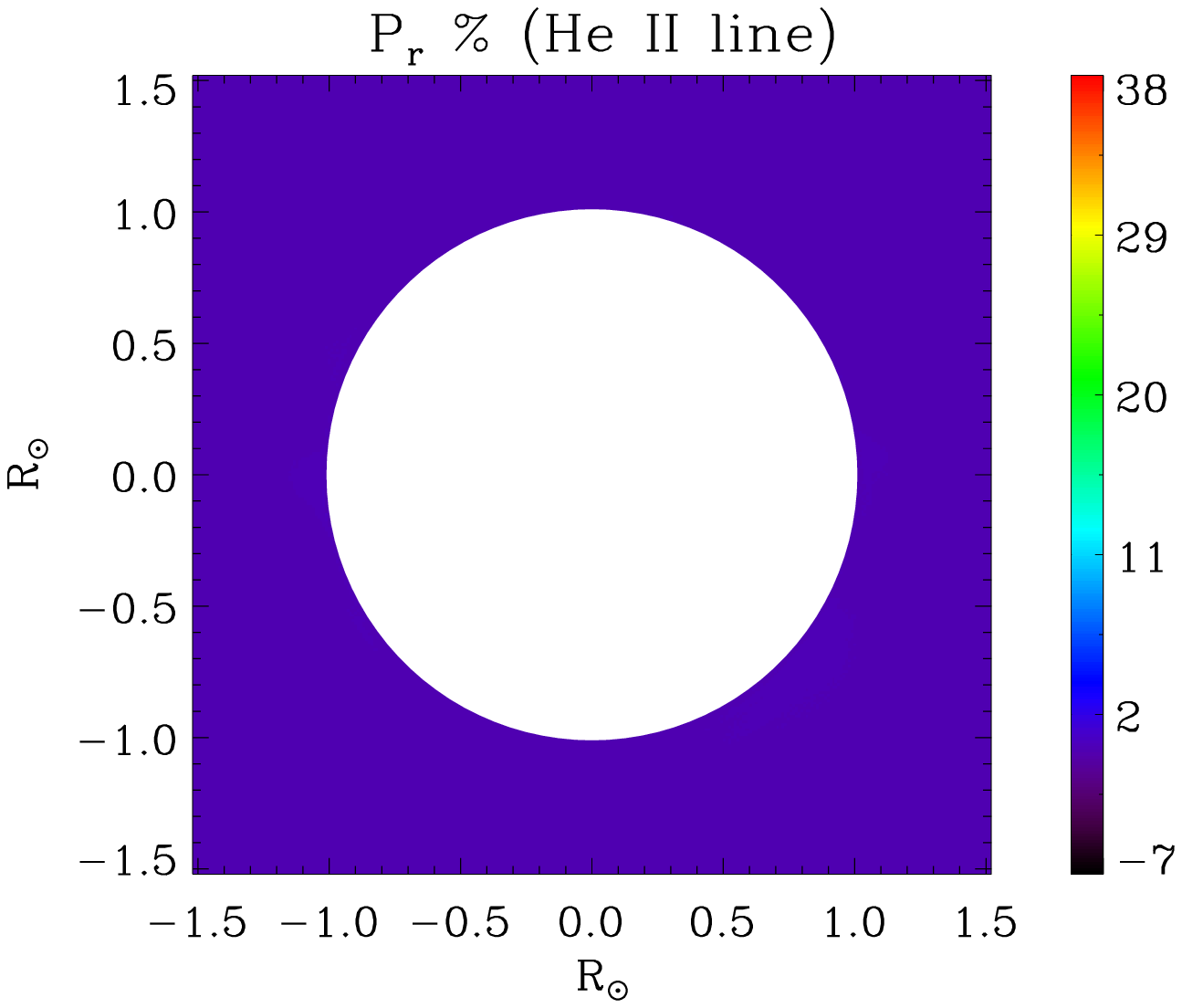}
	\includegraphics[scale=0.5]{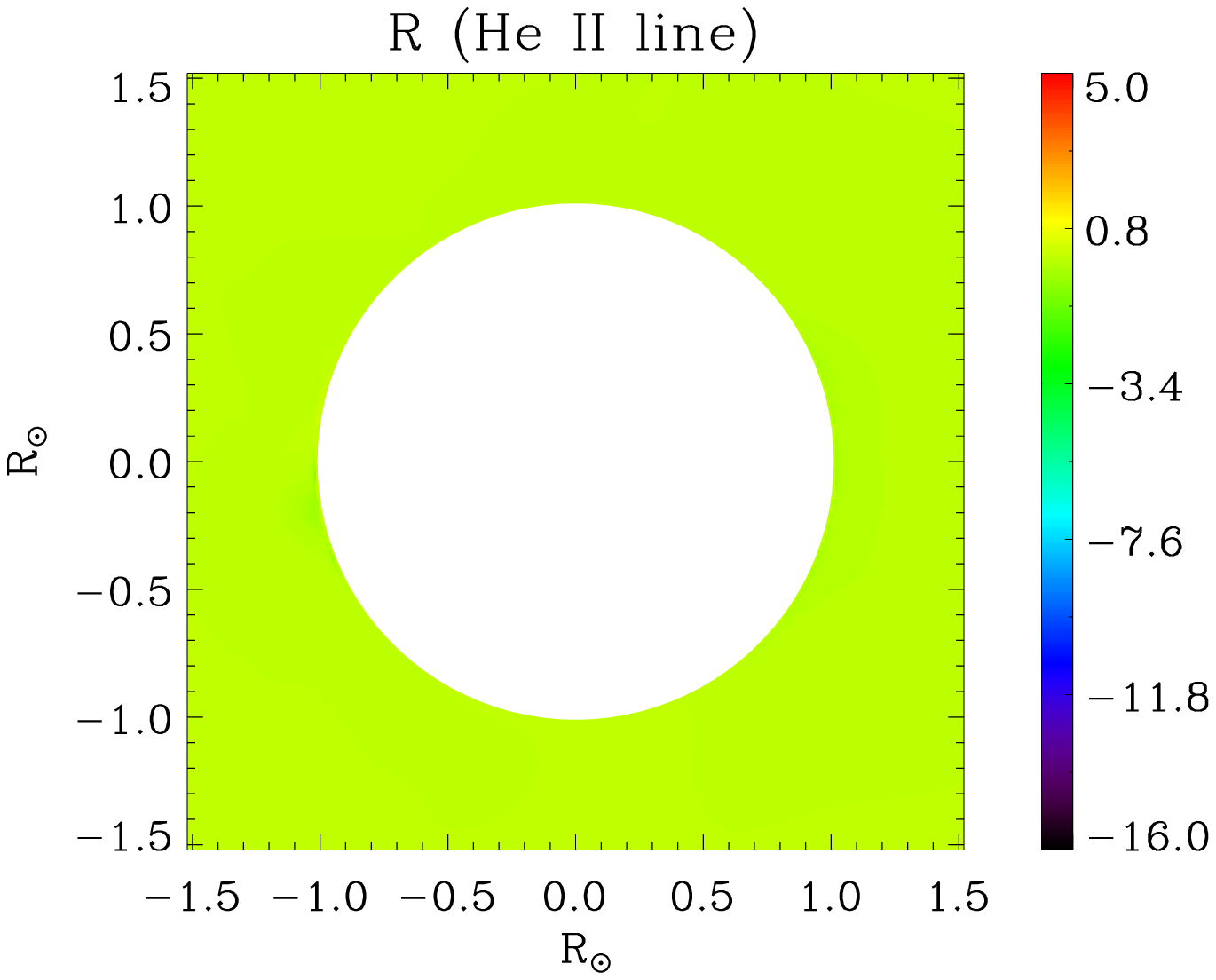}
	\caption{Same as Figure \ref{fig:cr2157-reldepolrot-SM} but for ``the dynamic model'' CR2138.}
	\label{fig:cr2138-reldepolrot-SM}
\end{figure*}

As seen in Figure \ref{fig:cr2138-param}, in ``the dynamic model'' the solar wind velocities are much larger, 
with values reaching 200 km/s even at coronal heights lower than 1.5R$_\odot$ 
(see also Figure \ref{fig:1dVandB}). A comparison of the number densities of \ion{H}{1} and \ion{He}{2} 
and of their corresponding Ly-$\alpha$ intensities in the two 
coronal models under consideration are given in Figure \ref{fig:number-density-photons}. 
The magnetic field of this model is weaker than in ``the magnetic model'' CR2157
previously considered, and thus the impact of the Hanle effect in the Ly-$\alpha$ line of \ion{H}{1} is smaller 
(see the top panels of Figure \ref{fig:cr2138-reldepolrot-SM}) and there is no hint of the 
Hanle effect in the Ly-$\alpha$ line of \ion{He}{2} (see the bottom panels of
Figure \ref{fig:cr2138-reldepolrot-SM}).

\begin{figure*}
	\includegraphics[scale=0.5]{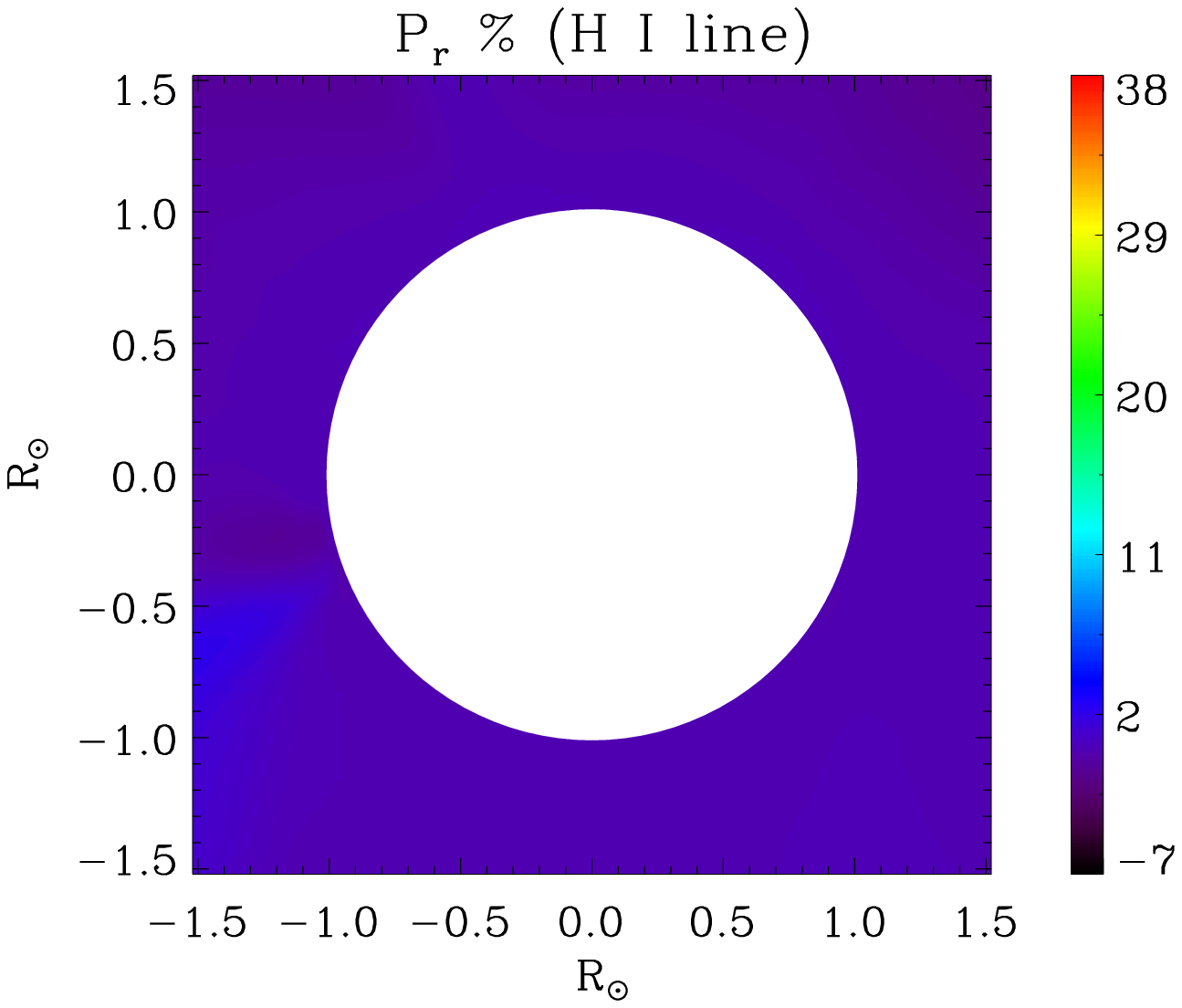}	
	\includegraphics[scale=0.5]{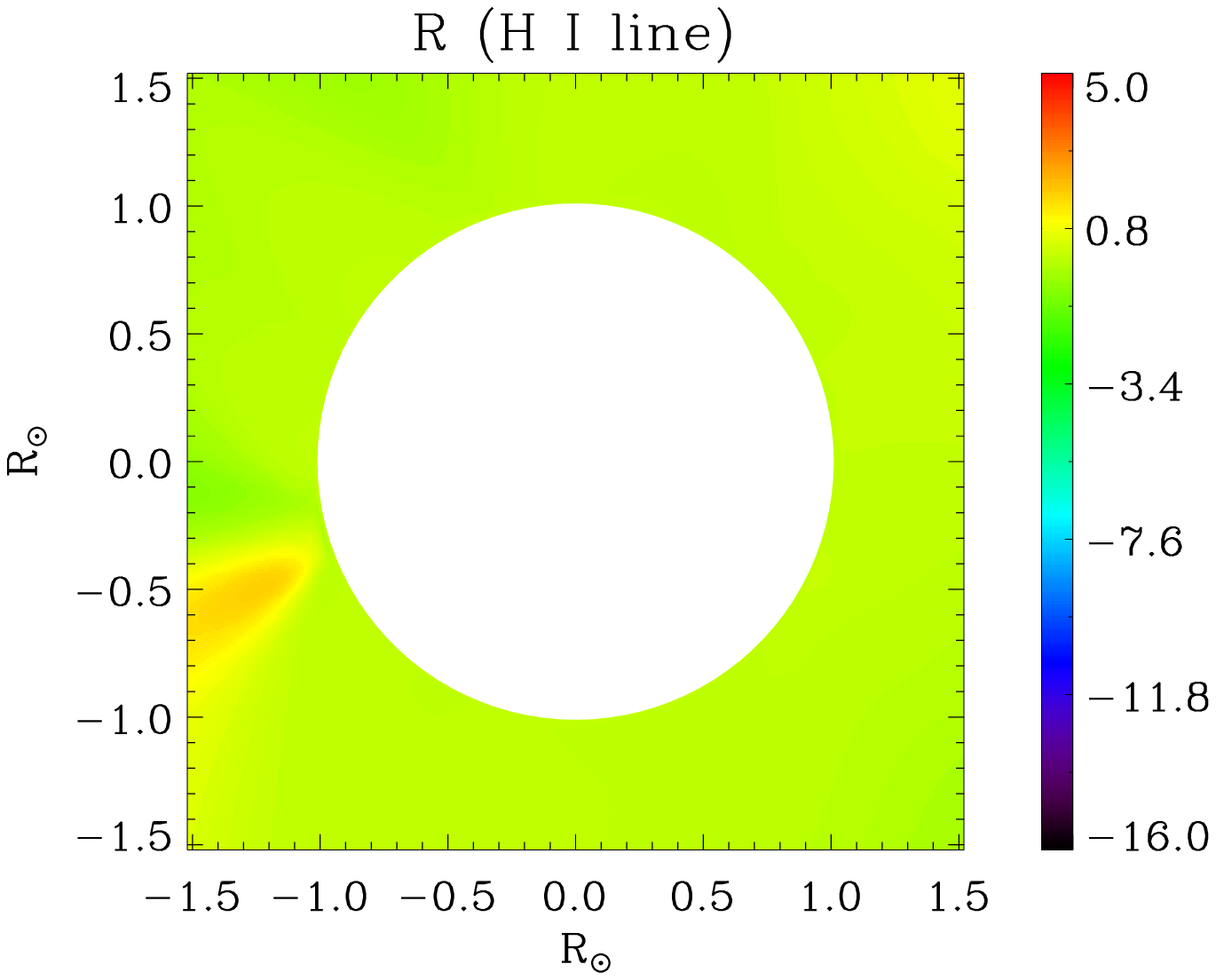}
	\includegraphics[scale=0.5]{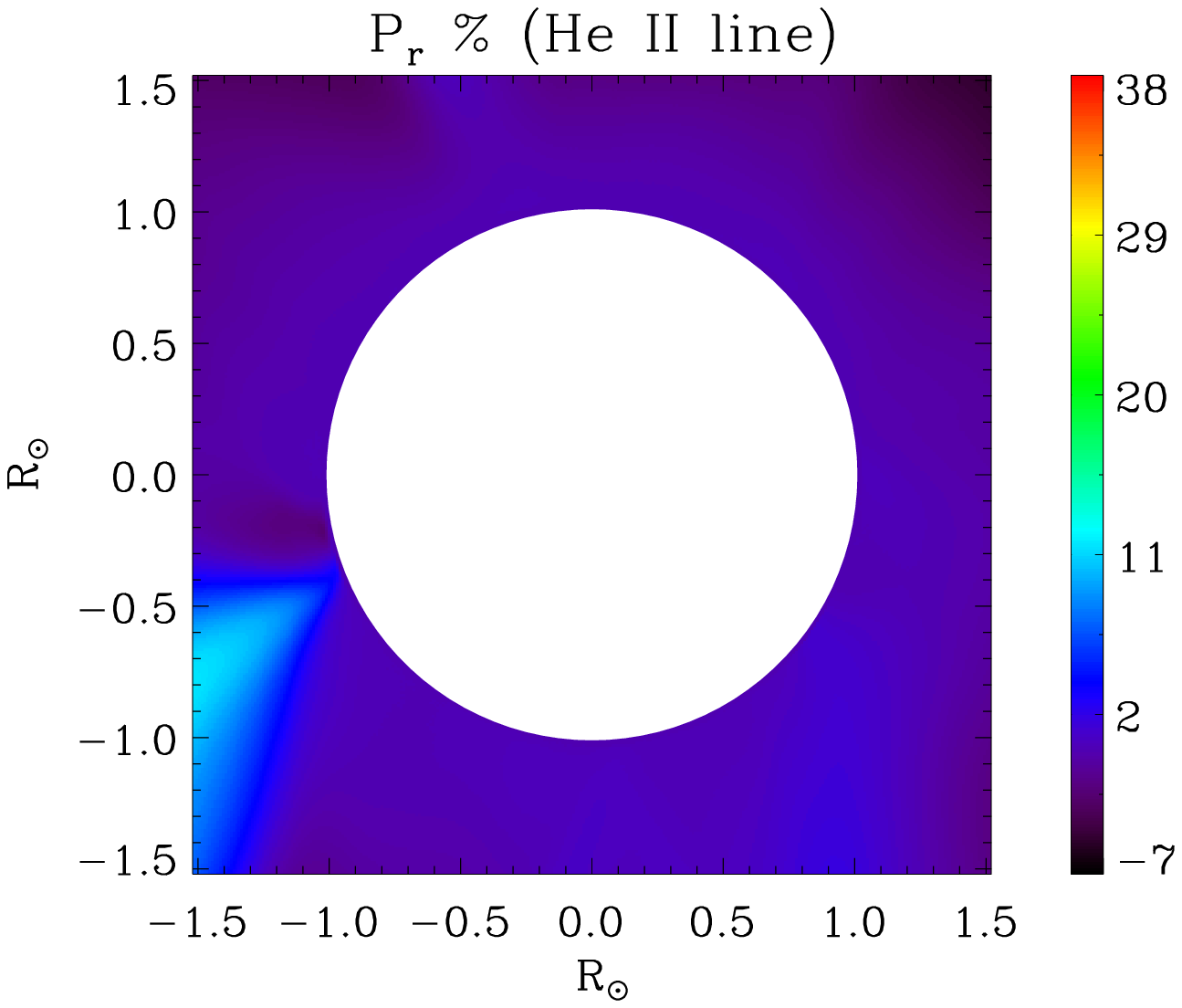}	
	\includegraphics[scale=0.5]{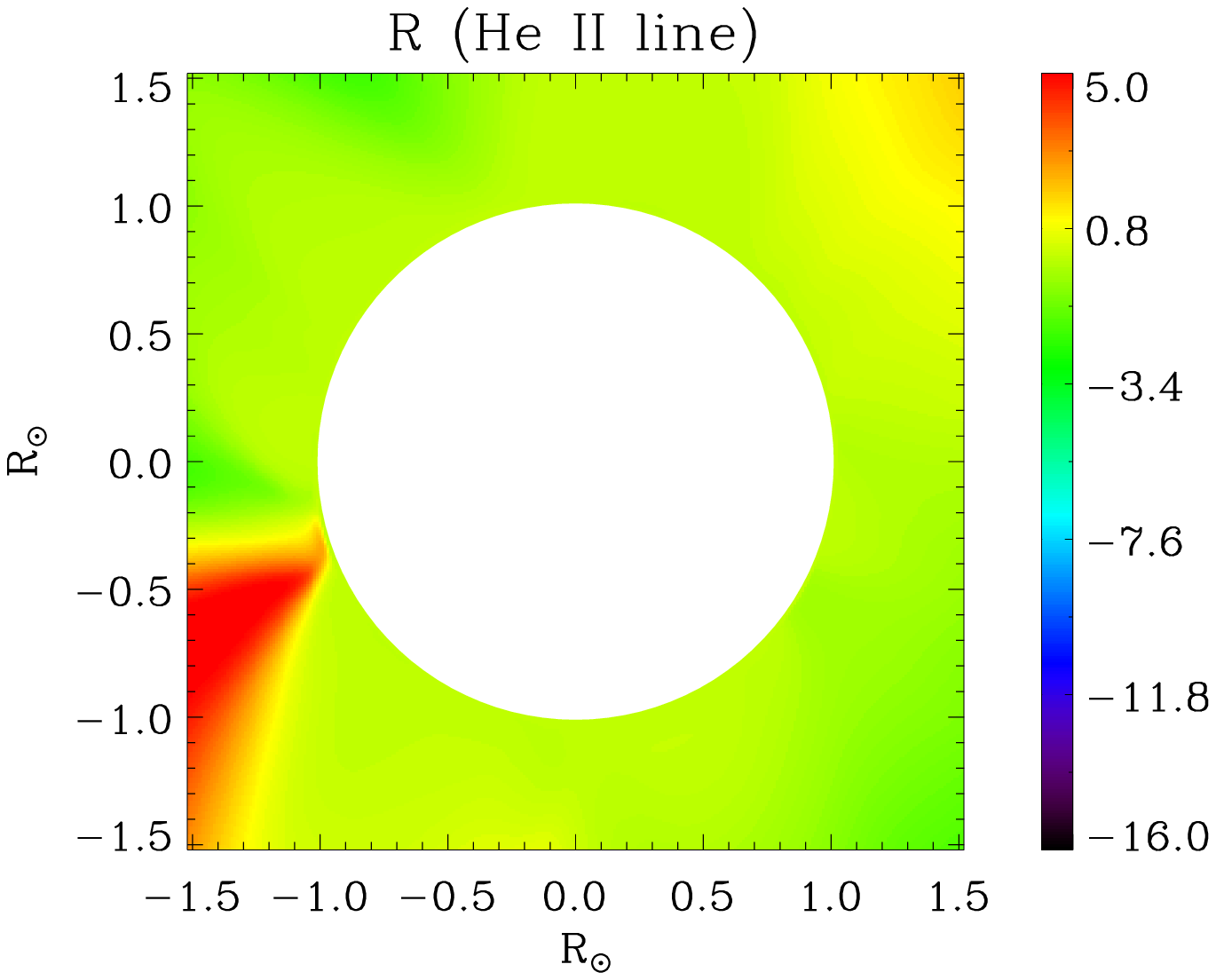}
	\caption{Same as Figure \ref{fig:cr2157-reldepolrot-SV} but for ``the dynamic model'' CR2138.}
	\label{fig:cr2138-reldepolrot-SV}
\end{figure*}

However, the sizable macroscopic velocities of ``the dynamic model'' wind produce an important depolarization 
and a rotation of the plane of linear polarization in the \ion{He}{2} line (see the bottom panels of 
Figure \ref{fig:cr2138-reldepolrot-SV}), much larger than in the broader Ly-$\alpha$ line of \ion{H}{1} 
(see the top panels of Figure \ref{fig:cr2138-reldepolrot-SV}). Interestingly, the solar wind velocities of 
``the dynamic model'' not only produce a strong depolarization and an anti-clockwise rotation of the 
polarization plane in the region of the model where the macroscopic velocity is more vigorous, but 
also an enhancement of the linear polarization and a clockwise rotation of the polarization plane 
in other regions of the coronal model where the macroscopic velocity is non-radial. 
In coronal regions similar to those in this model, the hydrogen 
Ly-$\alpha$ line can be considered to be a useful reference line for facilitating the detection of the fingerprints  
of the solar wind in the Ly-$\alpha$ line of \ion{He}{2}.

\section{Conclusions} \label{sec:conclusion}

The future of coronal spectropolarimetric diagnostics is through spectral lines with complementary 
sensitivities to the physical quantities of the mega-kelvin plasma. 
The linear polarization  
produced by the scattering of anisotropic radiation in suitably chosen spectral lines  
and its modification by the Hanle effect produced by the 
magnetic fields of the solar corona 
is one of the key mechanisms for obtaining empirical information on the 
solar corona. Here we have considered the 
Ly-$\alpha$ lines of \ion{H}{1} and \ion{He}{2} investigating their polarization 
in two 3D coronal models by Predictive Science Inc.
The reason for choosing these two lines is their very different 
sensitivities to the Hanle effect: the critical fields for the onset of the Hanle effect in the 
Ly-$\alpha$ lines of \ion{H}{1} and \ion{He}{2} being 53~G and 850~G, respectively. Therefore, for the 
field strengths expected for the solar corona the \ion{He}{2} Ly-$\alpha$ line
is practically insensitive to the 
Hanle effect, and here we have investigated whether we can use the Ly-$\alpha$ line 
of He {\sc ii} as a reference line 
for facilitating the determination of coronal magnetic fields
via the Hanle effect in the \ion{H}{1} Ly-$\alpha$ line.

However, there are other physical mechanisms 
such as the solar wind velocity, collisions, and active regions on 
the solar surface which affect the scattering polarization signals generated in the solar corona. 
Moreover, we have explored the impact of the solar wind velocities 
on the scattering polarization of the 
Ly-$\alpha$ lines of \ion{H}{1} and \ion{He}{2}. This is also important because 
the scattering atoms of \ion{H}{1} and \ion{He}{2} in the solar corona are irradiated 
by the spectral line radiation coming from the underlying solar disk, the intensity of which 
is an emission profile that is broader for the hydrogen Ly-$\alpha$ line. Therefore, the coronal 
atoms see a Doppler-shifted radiation field, which may have a significant impact on the 
anisotropy and symmetry properties of the radiation field seen by the coronal atoms.    
We have shown that, while the \ion{H}{1} line is mainly sensitive to the Hanle effect, the \ion{He}{2} line 
is mainly sensitive to the solar wind velocities. 
This is because of the much narrower \ion{He}{2} Ly-$\alpha$ 
line coming from the underlying atmosphere as compared to the \ion{H}{1} Ly-$\alpha$ line.
In the present investigation we have assumed that in the solar coronal models 
the helium abundance is uniform, but in a future investigation we plan to consider the possibility 
of significant spatial variations in the coronal helium abundance, as indicated by 
recent suborbital space measurements \citep{Moses+2020}.     

With the present technology, it should be possible to measure the intensity and polarization signals 
of both Ly-$\alpha$ lines within 0.5R$_\odot$ above the Sun's visible limb, but any measurement 
above these heights requires much larger integration times.
In regions of the solar corona
within 0.5R$_\odot$ of the solar surface, the solar wind velocity is usually low and the \ion{He}{2} line 
may be a useful reference line to determine the coronal magnetic field  
via the Hanle effect in the Ly-$\alpha$ line of \ion{H}{1} at 1216~\AA.
However, there might be dynamic events producing  
high solar wind velocities, even within 0.5R$_\odot$ 
above the limb. In such cases, the \ion{H}{1} Ly-$\alpha$ line 
may be a useful reference line to estimate the solar wind velocities via its effects on the 
\ion{He}{2} Ly-$\alpha$ line.

With the new diagnostic tool we have developed we 
can calculate the Stokes profiles of permitted and forbidden 
lines that emerge from 3D models of the solar corona and up to heliospheric distances. 
We are presently extending our theoretical investigations by considering various 
coronal forbidden lines, such as those to be observed with the  
Daniel K. Inouye Solar Telescope (DKIST).

\section*{Acknowledgements}
We acknowledge the funding received from the European Research Council (ERC) under the European Union's Horizon 2020 Research and Innovation Programme 
(ERC Advanced Grant agreement \mbox{No.~742265}), as well as through the projects PGC2018-095832-B-I00 and PGC2018-102108-B-I00 of the Spanish Ministry of Science, 
Innovation and Universities. This research made use of computing time available on the high-performance computing systems of the Instituto de Astrof\'\i sica de Canarias and 
we acknowledge the technical expertise and assistance provided by the Spanish Supercomputing Network. Thanks to the SOLARNET project, which has received funding from the European Union’s
Horizon 2020 Research and Innovation Programme under grant agreement
\mbox{No.~824135}, we obtained computing time 
at the Piz Daint supercomputer of the Swiss National
Supercomputing Centre (CSCS).
CHIANTI is a collaborative project involving George Mason University, the University of 
Michigan (USA), University of Cambridge (UK) 
and NASA Goddard Space Flight Center (USA).

\bibliography{dayananda}{}
\bibliographystyle{aasjournal}
 

\end{document}